\documentclass{amsart}
\usepackage{graphicx,amssymb}
\newtheorem{thm}{Theorem}[section]
\newtheorem{cor}[thm]{Corollary}

\theoremstyle{definition}
\newtheorem{defn}[thm]{Definition}
\theoremstyle{remark}

\numberwithin{equation}{section}

\newcommand{\mbold}[1]{\mbox{\boldmath${#1}$}}
\def\beq{\begin{eqnarray}}
\def\eeq{\end{eqnarray}}
\begin{document}

\title[A dynamical systems approach to the tilted Bianchi models]{A dynamical systems approach to the tilted Bianchi models of solvable type}%
\author{Alan Coley, Sigbj{\o}rn Hervik}%
\address{Department of Mathematics \& Statistics, Dalhousie University,
Halifax, Nova Scotia,
Canada B3H 3J5}%
\email{aac@mathstat.dal.ca, herviks@mathstat.dal.ca}%


\date{\today}%
\begin{abstract}

We use a dynamical systems approach to analyse the tilting
spatially homogeneous Bianchi models of solvable type 
(e.g., types VI$_h$  and VII$_h$)  
with a perfect fluid and a linear barotropic $\gamma$-law equation of
state. In particular, we  study the late-time behaviour of
 tilted Bianchi models, with an emphasis on the existence
of equilibrium points and their stability properties. We briefly discuss the tilting Bianchi type V models and the late-time asymptotic behaviour of irrotational Bianchi VII$_0$ models.   We prove the
important result that for non-inflationary Bianchi type VII$_h$
models vacuum plane-wave
solutions  are the only future attracting
equilibrium points in the Bianchi type VII$_h$ invariant set. We
then investigate the dynamics close to the plane-wave solutions in
more detail, and discover some new features that arise in the
dynamical behaviour of Bianchi cosmologies with the inclusion of
tilt. We point out that in a tiny open set of parameter space in the type
IV model (the loophole) there exists closed curves which act as
attracting limit cycles. More interestingly,
in the Bianchi type VII$_h$ models there is a bifurcation in which
a set of equilibrium points turn into closed orbits. There is a
region in which both sets of closed curves coexist, and it appears that for the type VII$_h$ models in this region  
the solution curves approach a compact surface which is topologically
a torus.

\end{abstract}
\maketitle

\section{Introduction}

Spatially homogeneous (SH) models are very important in
contemporary cosmology for two main reasons. First, these models
are useful in  analyzing the physical effects of the Universe
which are affected by anisotropy in the rate of expansion. Second,
SH Bianchi universes are relatively straightforward to analyse
since the Einstein field equations for these cosmologies reduce to
ordinary differential equations (DEs).

A SH cosmology is said to be {\it tilted} if the fluid velocity
vector is not orthogonal to the group orbits, otherwise the model
is said to be {\it non-tilted} \cite{KingEllis}. Tilted universes
are of  considerable interest since they contain many features of
physical relevance, including rotation, the conditions needed for
the appearance of closed time-like curves and some properties of
quantum cosmologies. In this paper we shall assume that the matter
content of the universe is a perfect fluid with equation of state
$p = (\gamma -1) \mu$, where $\gamma$ is constant, which includes
the important cases of dust and radiation, $\gamma =1$ and $\gamma
= \frac{4}{3}$, respectively.

Much work has been done in analyzing non-tilting SH models
\cite{EM,DS1,DS2}. However, less is known about the tilted models.
Since Bianchi models with a \emph{tilted} fluid \cite{KingEllis}
contain up to three additional degrees of freedom, it is evident
that there is an increase in dynamical complexity and new features
emerge with the inclusion of tilt.

For the class of tilted SH models, the Einstein field equations
have been written as an autonomous DE in a number of different
ways \cite{BKL,BS,rosjan,BN,BS1}, but due to the complexity of the
equations a detailed analysis of the dynamics is difficult. A
subclass of models of Bianchi type V
\cite{Shikin,Collins,CollinsEllis,HWV} and models of Bianchi type
II \cite{HBWII} have been studied using dynamical systems theory.
In particular, a complete description of the dynamics of tilted SH
cosmologies of Bianchi type II  was given in \cite{HBWII}, with
particular emphasis on the qualitative analysis of the dynamics
near the initial singularity and at late times. It was found that
for $\gamma < 2$, the tilt destabilizes the Kasner solutions
leading to a Mixmaster-like initial singularity, with the tilt
being dynamically significant, while at late times the tilt
becomes dynamically negligible unless $\gamma > \frac{10}{7}$.  It
was also found that the tilt does not destabilize the flat
Friedmann model, with the result that the presence of tilt
increases the likelihood of intermediate isotropization,
supporting earlier results found in tilted Bianchi V models
\cite{HWV}.

Recently more general tilting Bianchi type universes have been
studied, with particular emphasis on exact solutions of the
Einstein equations which are attractors for more general classes
of solution. In \cite{BHtilted} a local stability analysis of a
wide range of tilting SH universes, including Bianchi class B
models and the type VI$_{0}$ model of class A, were studied to
elucidate their late-time behaviours. The asymptotic late-time
behaviour of general tilted Bianchi type VI$_0$ universes was
analysed in detail in \cite{hervik}, and this was generalized to
Bianchi type VI$_0$ cosmological models containing two tilted
$\gamma$-law perfect fluids in \cite{coleyhervik}. All of the
future stable equilibrium points for various subclasses of tilted
type VI$_0$ models, as well as for the general tilted type VI$_0$
models, were found  (and all of the bifurcation values identified;
e.g., $\gamma = 2/3$, $\gamma = 10/9$, $\gamma =6/5$ and $\gamma
=4/3$).

In this paper we will generalize previous work and use a dynamical
systems approach  to analyse the tilting Bianchi type VI$_h$  and
VII$_h$ models (and their subclasses) containing a $\gamma$-law
perfect fluid. We shall adopt the formalism employed in
\cite{hervik}. These Bianchi models are of general measure in the
space of all SH models and are sufficiently general to account for
many interesting phenomena. In particular, the Bianchi type
VII$_h$ models are the most general models that contain the open
Friedmann-Robertson-Walker (FRW) models and are consequently of special
interest.

The paper is organised as follows. In the next section we shall
write down the equations of motion using the orthonormal frame
formalism and discuss the  invariant subspaces. Various
gauge choices are considered, and it is emphasised that the
'N-gauge' is useful for studying the stability of the plane wave
solutions. In the next section we briefly discuss models of type V
and VII$_0$. In particular we show that the self-similarity breaking which occurs in the non-tilted type VII$_0$, is also present for tilted, irrotational, type VII$_0$ models. 

We then study the late-time behaviour of various tilted Bianchi
models, with an emphasis on the existence of equilibrium points
and their stability properties. In particular, we study the
stability properties of the vacuum plane-wave spacetimes, and show
that for $2/3<\gamma<2$ there will always be future stable
plane-wave solutions in the set of type IV and VII$_h$ tilted
Bianchi models.  Indeed, we prove a theorem concerning the future
evolution in the invariant type V and type VII$_h$ sets, and in
the case of Bianchi type VII$_h$ models it is shown that the only
future attracting equilibrium points for non-inflationary fluids
($\gamma>2/3$) are the plane-wave solutions. We also study the
form of the asymptotic tilt velocity for these attractors. For inflationary type  fluids ($0<\gamma<2/3$) the situation is simpler; we show that the cosmic no-hair theorem is valid in this case. 

We then study the dynamics in the neighbourhood of the vacuum
plane-wave solutions. In the type IV models, we discover that
there is a tiny region of the parameter space in which there
exists a closed orbit, which acts as an attracting limit cycle.
Moreover, for the type VII$_h$ models 
more attracting closed orbits appear. 
A numerical analysis which is presented in a companion paper \cite{HHC}, shows that there is an open set of
parameter space such that the solution curves approach a compact
surface which is topologically a torus in the type VII$_h$ models.

We also study the stability of the plane waves in Bianchi type VI$_h$ models
and the stability of the Collinson-French vacuum solution in the
exceptional Bianchi type VI$_{-1/9}$ models. In the final section we
summarize our results, discuss possible early time behaviour and
briefly describe future work. Some more technical results are
presented in the appendices.

Finally, as stressed in \cite{DS1,DS2}, an important mathematical
link exists between the various classes in the {\it state space}
hierarchy. The physical state of a cosmological model at an
instant of time is represented by a point in the state space,
which is finite dimensional for SH models and infinite dimensional
otherwise. This structure opens the possibility that the evolution
of a model in one class may be approximated, over some time
interval, by a model in a more special class. Thus it is plausible
that understanding the dynamics at one level of complexity,  such
as for example the tilting Bianchi type VI$_h$  and VII$_h$
models, will shed light on the possible dynamical behaviour at a
higher level, such as special classes of inhomogeneous
cosmological models.

\section{Equations of motion}
Let us  consider the Bianchi models
containing a tilted $\gamma$-law perfect fluid. We will investigate the  models having an Abelian $G_2$ subgroup. For the Bianchi models, these correspond to the Lie algebras, $\mathcal{A}$, where $\mathcal{A}$ is \emph{solvable}\footnote{A Lie algebra, ${\mathfrak g}$, is called \emph{solvable} if  its derived series, defined as ${\mathfrak g}_0=\mathfrak{g}$, ${\mathfrak g}_i=[{\mathfrak g}_{i-1},{\mathfrak g}_{i-1}]$, terminates; i.e., there exists a $k$ such that ${\mathfrak g}_k=0$.}; i.e., types I-VII$_h$. Types VIII and IX are both semi-simple and have no Abelian $G_2$ subgroup.
For the Bianchi cosmologies -- which admit a simply
transitive symmetry group acting on the spatial hypersurfaces -- we can
always write the line-element as 
\[
{\rm d}s^{2}=-{\rm d}t^{2}+\delta_{ab}{\mbox{\boldmath${\omega}$}}^{a}{\mbox{\boldmath${%
\omega}$}}^{b},
\]%
where ${\mbox{\boldmath${\omega}$}}^{a}$ is a triad of one-forms
obeying 
\[
\mathbf{d}{\mbox{\boldmath${\omega}$}}^{a}=-\frac{1}{2}C_{~bc}^{a}{%
\mbox{\boldmath${\omega}$}}^{b}\wedge {\mbox{\boldmath${\omega}$}}^{c},
\]%
and $C_{~bc}^{a}$ depend only on time and are the structure constants
of the Bianchi group type 
under consideration. The structure constants $C_{~bc}^{a}$ can be split into
a vector part $a_{b}$, and a trace-free part $n^{ab}$ by \cite{EM} 
\[
C_{~bc}^{a}=\varepsilon _{bcd}n^{da}-\delta _{~b}^{a}a_{c}+\delta
_{~c}^{a}a_{b}.
\]%
The matrix $n^{ab}$ is symmetric, and, using the Jacobi identity, $%
a_{b}=(1/2)C_{~ba}^{a}$ is in the kernel of $n^{ab}$ 
\[
n^{ab}a_{b}=0.
\]
The time-like vector ${\bf e}_0$ is chosen orthogonal to the group orbits and is given by ${\bf e}_0=\partial/\partial t$ where $t$ is the cosmological time. It is  necessary to introduce a dimensionless time variable, $\tau$, defined by
\beq
\frac{{\rm d} t}{{\rm d} \tau}=\frac{1}{H},
\eeq
where $H$ is the Hubble scalar. 
Following \cite{DS1,DS2} we introduce expansion-normalised variables and the equations of motion can be given as an autonomous system of DEs (see \cite{vanElst}).

The equations of
motion will in general depend on the choice of gauge. We will choose an
orthonormal frame where ${\bf e}_1$ points in the direction of
the vector $a_{b}=\frac 12C^{a}_{~ba}$, but we will
leave the remaining frame vectors ${\bf e}_2$ and ${\bf e}_3$
defined up to a rotation. Explicitly, this frame rotation is
given by
\beq \tilde{\bf e}_2=
\cos\phi {\bf e}_2+\sin\phi{\bf e}_3, \quad 
\tilde{\bf e}_3= -\sin\phi {\bf e}_2+\cos\phi{\bf e}_3,
\label{eq:framerot}
 \eeq
where $\phi$ is a real function.
This frame rotation defines the remaining gauge freedom we have
in our system of equations.

Following the notation in \cite{vanElst}, the expansion-normalised variables are now
\beq
\Sigma_{ab}&=&\begin{bmatrix}
-2\Sigma_+ & \sqrt{3}\Sigma_{12} & \sqrt{3}\Sigma_{13} \\
\sqrt{3}\Sigma_{12} & \Sigma_++\sqrt{3}\Sigma_- &
\sqrt{3}\Sigma_{23}
\\
 \sqrt{3}\Sigma_{13} & \sqrt{3}\Sigma_{23} & \Sigma_+-\sqrt{3}\Sigma_-
\end{bmatrix}, \nonumber \\
N_{ab}&=&\sqrt{3}\begin{bmatrix} 0 & 0 & 0 \\
0 & \bar{N}+N_- & N_{23} \\
0 & N_{23} & \bar{N}-N_-
\end{bmatrix}.
\eeq
To facilitate the remaining gauge freedom we will utilize complex
variables (which will be written in bold typeface)
: \beq {\bf
N}_{\times}&=&N_-+iN_{23}, \quad
{\mbold\Sigma}_{\times}=\Sigma_-+i\Sigma_{23}\nonumber \\
{\mbold\Sigma}_1&=&\Sigma_{12}+i\Sigma_{13}, \quad
{\bf v}=v_2+iv_3. \eeq
A gauge transformation is then given by
\beq
\left({\bf N}_{\times},{\mbold\Sigma}_{\times},{\mbold\Sigma}_1,{\bf v}\right)&\mapsto& \left(e^{2i\phi}{\bf N}_{\times},e^{2i\phi}{\mbold\Sigma}_{\times},e^{i\phi}{\mbold\Sigma}_1,e^{i\phi}{\bf v}\right).
 \eeq
Here, the phase $\phi$ can be a function of time and corresponds to
a rotation of frame according to  eq.(\ref{eq:framerot}). From the
above we see that the variables ${\bf N}$ and
${\mbold\Sigma}_{\times}$ are spin-2 objects and
${\mbold\Sigma}_1$ and ${\bf v}$ are spin-1 objects under the
frame rotations. The complex conjugates, which we will denote with
an asterisk, transform similarly: \beq
\left({\bf N}_{\times}^*,{\mbold\Sigma}_{\times}^*,{\mbold\Sigma}_1^*,{\bf v}^*\right)&\mapsto& \left(e^{-2i\phi}{\bf N}_{\times}^*,e^{-2i\phi}{\mbold\Sigma}_{\times}^*,e^{-i\phi}{\mbold\Sigma}_1^*,e^{-i\phi}{\bf v}^*\right).
\eeq

 In the
following, we will explicitly leave the gauge function $\phi$ in
the equations of motion and we will assume that $\phi$ is given
with respect to a frame for which $\phi'=0$ implies $R_1=0$ where $R_a$ is the (expansion-normalised) local angular velocity of a Fermi-propagated axis with respect to the triad ${\bf e}_a$. Thus
we will replace the function $R_1$ with the function $\phi$.
Explicitly, we have set $R_1=-\phi'$. At this stage it is
important to note that all the physical variables have to be
independent of the gauge. Hence, the objects ${\bf N}_{\times}$
and ${\bf v}$ are themselves not physical variables; only scalars
constructed from the above are gauge independent (we will define
what we mean by scalars below).

 For the expansion-normalised variables the equations of motion
are:
\beq \Sigma_+'&=&
(q-2)\Sigma_++{3}|{\mbold\Sigma}_1|^2-2|{\bf N}_{\times}|^2
+\frac{\gamma\Omega}{2G_+}\left(-2v_1^2+|{\bf v}|^2\right) \label{eq:Sigma+eq}\\
{\mbold\Sigma}_{\times}'&=&
(q-2+2i\phi'){\mbold\Sigma}_{\times}+\sqrt{3}{\mbold\Sigma}_{1}^2
-2{\bf N}_{\times}(iA+\sqrt{3}\bar{N})
+\frac{\sqrt{3}\gamma\Omega}{2G_+}{\bf v}^2
\\
{\mbold\Sigma}'_{1}&=& \left(q-2-3\Sigma_++i\phi'\right){\mbold
\Sigma}_{1} -\sqrt{3}{\mbold\Sigma}_{\times}{\mbold\Sigma}_{1}^*
+\frac{\sqrt{3}\gamma\Omega v_1}{G_+}{\bf v}
\\
 {\bf N}_{\times}'&=& \left(q+2\Sigma_++2i\phi'\right){\bf N}_{\times}+2\sqrt{3}{\mbold\Sigma}_{\times}\bar{N}\\
\bar{N}'&=&
\left(q+2\Sigma_+\right)\bar{N}+2\sqrt{3}\mathrm{Re}\left({\mbold\Sigma}_{\times}^*{\bf
N}_{\times}\right)
\\
A'&=& (q+2\Sigma_+)A . \label{eq:Aeq}\eeq
The equations for the fluid are
\beq
\quad \Omega'&=& \frac{\Omega}{G_+}\Big\{2q-(3\gamma-2)+2\gamma
Av_1
 +\left[2q(\gamma-1)-(2-\gamma)-\gamma\mathcal{S}\right]V^2\Big\}\label{eq:Omega}
 \quad \\
 v_1' &=& \left(T+2\Sigma_+\right)v_1-2\sqrt{3}\mathrm{Re}\left({\mbold\Sigma}_{1}{\bf v}^*\right) -A|{\bf v}|^2+\sqrt{3}\mathrm{Im}({\bf N}_{\times}^*{\bf v}^2)\\
 {\bf v}'&=& \left(T-\Sigma_++i\phi'+Av_1-i\sqrt{3}\bar{N}v_1\right){\bf v}-\sqrt{3}\left({\mbold\Sigma}_{\times}+i{\bf N}_{\times}v_1\right){\bf v}^* \\
  V'&=&
\frac{V(1-V^2)}{1-(\gamma-1)V^2}\left[(3\gamma-4)-2(\gamma-1)Av_1-\mathcal{S}\right]
\eeq where
\beq q&=& 2\Sigma^2+\frac
12\frac{(3\gamma-2)+(2-\gamma)V^2}{1+(\gamma-1)V^2}\Omega\nonumber \\
\Sigma^2 &=& \Sigma_+^2+|{\mbold{\Sigma}}_{\times}|^2+|{\mbold\Sigma}_{1}|^2\nonumber \\
\mathcal{S} &=& \Sigma_{ab}c^ac^b, \quad c^ac_{a}=1, \quad v^a=Vc^a,\quad \nonumber \\
 V^2 &=& v_1^2+|{\bf v}|^2,\quad  \nonumber \\
G_+ &=& 1+(\gamma-1)V^2, \nonumber \\
 T&=& \frac{\left[(3\gamma-4)-2(\gamma-1)Av_1\right](1-V^2)+(2-\gamma)V^2\mathcal{S}}{1-(\gamma-1)V^2}.
\label{eq:defs}
\eeq
These variables are subject to the constraints
\beq
1&=& \Sigma^2+A^2+|{\bf N}_{\times}|^2+\Omega \label{const:H}\\
0 &=& 2\Sigma_+A+2\mathrm{Im}({\mbold\Sigma}_{\times}^*{\bf N}_{\times})+\frac{\gamma\Omega v_1}{G_+} \label{const:v1}\\
0 &=&
\mbold{\Sigma}_{1}(i\bar{N}-\sqrt{3}A)+i{\mbold{\Sigma}}^*_{1}{\bf
N}_{\times}+\frac{\gamma\Omega {\bf v}}{G_+} \label{const:v2}
 \\
0&=& A^2+3h\left(|{\bf
N}_{\times}|^2-\bar{N}^2\right)\label{const:classB} \eeq The
parameter $\gamma$ will be assumed to be in the interval
$\gamma\in ( 0,2)$.

\section{The state space}
The state vector can be considered as ${\sf
X}=(\Sigma_+,{\mbold\Sigma}_{\times},{\mbold\Sigma}_1,{\bf
N}_{\times},\bar{N},A,v_1,{\bf v})$ modulo the constraints eqs.
(\ref{const:v1}), (\ref{const:v2}) and  (\ref{const:classB}). The
latter is the group constraint, which we assume determines the
parameter $h$ and consequently is not a free variable. The Bianchi
identities ensure that the constraints
(\ref{const:H})-(\ref{const:v2}) are first integrals; thus
satisfying the constraints on a initial hypersurface is sufficient
to ensure that they are satisfied at all times (see also the
Appendix where the expressions for their time derivatives are
explicitly given).

The variable $\Omega$ is determined from the Hamiltonian
constraint, eq.(\ref{const:H}),  and can thus be eliminated. The
tuple ${\sf X}$ is 12-dimensional, but the constraints reduce the
number of independent components to 8.  In the above equations the
function $\phi$ carries the choice of gauge and does not have an
evolution equation. Specifying this function determines the gauge
completely. We can use this function to eliminate one degree of
freedom. This reduces the dimension of the state space to 7 (for a
given $h$); i.e., the physical state space is 7-dimensional. Again
we note that all of the physical information can be extracted from
considering real scalars.

Let us list some relevant invariant subspaces necessary for this discussion.
\begin{enumerate}
\item{$T(\mathcal{A})$} The full state space of tilted solvable Bianchi models.
\item{$T(VI_h)$} Type VI$_h$: $|{\bf N}_{\times}|^2-\bar{N}^2>0$.
\item{$T(VII_h)$} Type VII$_h$: $|{\bf N}_{\times}|^2-\bar{N}^2<0$.
\item{$T(VI_0)$} Type VI$_0$: $|{\bf N}_{\times}|^2-\bar{N}^2>0$, $A=0$.
\item{$T(VII_0)$} Type VII$_0$: $|{\bf N}_{\times}|^2-\bar{N}^2<0$, $A=0$.
\item{$T(V)$} Type V: $|{\bf N}_{\times}|=\bar{N}=0, A\neq 0$.
\item{$T(IV)$} Type IV: $|{\bf N}_{\times}|^2-\bar{N}^2=0$, $A\neq 0$.
\item{$T(II)$} Type II: $|{\bf N}_{\times}|^2-\bar{N}^2=0$, $A=0$.
\item{$B(I)$} Type I: $|{\bf N}_{\times}|=\bar{N}=A=0$.
\item{$T_1(\mathcal{A})$}Non-exceptional, one-component tilted fluid: ${\mbold\Sigma}_1={\bf v}=0$.
\item{$B(\mathcal{A})$} Non-tilted, non-exceptional: ${\mbold\Sigma}_1={\bf v}=v_1=0$.
\item{$\partial T(\mathcal{A})$} ``Tilted'' vacuum boundary: $\Omega=0$.
\item{$\partial B(\mathcal{A})$} Non-tilted vacuum boundary: $\Omega={\bf v}=v_1=0$.
\end{enumerate}
Note that the Hamiltonian constraint, ensures that 
\beq
\Sigma_+^2+|{\mbold\Sigma}_{\times}|^2+|{\mbold\Sigma}_1|^2+|{\bf
N}_{\times}|^2+ A^2 \leq 1. 
\eeq 
In addition we will require that
the tilt velocities are not superluminal; i.e. 
\beq v_1^2+|{\bf v}|^2\leq 1. 
\eeq 
The variable $\bar{N}$ may or may not be bounded.
However, from eq.(\ref{const:classB}) we see that $\bar{N}$ can
only be unbounded for $T(VII_0)$. Hence, for all models except
type VII$_0$, the state space is compact.

 \subsection{Equilibrium points and gauge independent quantities}
\begin{defn}[Scalar]
We will call a quantity which is independent of the
transformation, eq. (\ref{eq:framerot}), a scalar. Scalars are
thus \emph{gauge independent quantities}.
\end{defn}
Examples of such scalars are easy to construct. Note that
$\Sigma_+$, $\bar{N}$ and $A$ are all scalars. Furthermore,
objects like ${\bf N}_{\times}^*{\bf N}_{\times}$, ${\bf
N}_{\times}^*{\bf v}^2$, ${\mbold\Sigma}_1{\bf v}^*$, are also
scalars. In the above formalism it is also necessary to provide  a
gauge independent definition of equilibrium points.
\begin{defn}[Equilibrium points]
A set $P$ is said to be (an) equilibrium point(s) if all scalars are
constants on $P$ as functions of $\tau$.
\end{defn}
We note that if different points in $P$  have different scalars
then the points may represent different physical configurations.
The constancy as a function of $\tau$ is then the usual
requirement for equilibrium points.

For the non-tilted, non-exceptional models, ${\mbold\Sigma_1}={\bf
v}=0$, and thus from the gauge dependent quantities
${\mbold\Sigma_{\times}}$ and ${\bf N}_{\times}$ there are only
four real scalars.  These are related via the identity \beq
\left|{\bf
N}_{\times}\right|^2\left|{\mbold\Sigma}_{\times}\right|^2=\left[{\rm
Re}\left({\bf
N}_{\times}{\mbold\Sigma}^*_{\times}\right)\right]^2+\left[{\rm
Im}\left({\bf
N}_{\times}{\mbold\Sigma}^*_{\times}\right)\right]^2.
\label{eq:NSconstraint}\eeq Hence, there are only three
independent scalars constructed from ${\mbold\Sigma_{\times}}$ and
${\bf N}_{\times}$. In the non-tilted analysis, Hewitt and
Wainwright \cite{HWClassB} used these scalars as dynamical
variables (instead of ${\mbold\Sigma_{\times}}$ and ${\bf
N}_{\times}$ themselves) at the cost of obtaining the additional
constraint (\ref{eq:NSconstraint}). In principle, we could follow
the same approach for the tilted analysis. This choice might be appropriate for certain subspaces (like $T_1(\mathcal{A})$); however, in general, this would
lead to too many constraints (similar to eq.(\ref{eq:NSconstraint}))
to be useful in practice.

\subsection{Some Choices of Gauge}There are different
convenient choices of gauges for the different invariant
subspaces. Here we will discuss three such choices which are
useful for different purposes.

\subsubsection{"F-gauge": $\phi'=0$}
The simplest possible choice of $\phi$ is to let $\phi'=0$. Note
that this does not specify the gauge completely; we still have a
constant $U(1)$ gauge freedom left. However, the F-gauge is
convenient for several reasons. First, the system of equations are
defined everywhere on the state space. Second, the equations of
motion reduce to an autonomous system of equations which define a
dynamical system (with 8 variables). However, in this gauge there
is still a constant $U(1)$-transformation remaining which can, for
example, be used for simplifying the initial data. The initial
data will then, modulo this constant transformation, define a
unique path in the state space.

It should also be noted that in this gauge the equilibrium points
of the dynamical system do not necessarily have ${\sf X}'=0$.
Hence, for calculating equilibrium points there might be a more
convenient gauge choice.

\subsubsection{"${\bf N}$-gauge": ${\bf N}_{\times}$ purely imaginary}
This choice of gauge uses the gauge-function $\phi$ to simplify
the function ${\bf N}_{\times}$. We can, for example, choose the
function ${\bf N}_{\times}$ to be purely imaginary. To ensure that
${\bf N}_{\times}$ is purely imaginary at all times we have to
choose $\phi'=\sqrt{3}\lambda\Sigma_-$, where $\lambda$ is defined
by 
\beq
\bar{N}=\lambda\mathrm{Im}({\bf N}_{\times}). 
\eeq
The equation
for $\bar{N}$ can then be replaced with an equation for $\lambda$
which ensures a closed system of equations.
The advantages of this gauge is that there is no gauge freedom
left and the equilibrium points are given by ${\sf X}'=0$.
However,  care is needed because the parameter $\lambda$ is
ill-defined near ${\bf N}_{\times}=0$. Close to ${\bf
N}_{\times}=0$ the parameter $\lambda$ will diverge and the
dynamical system will cease to be well defined. Note, however,
that in $T(VI_h)$ this cannot happen since $\lambda^2<1$ in
$T(VI_h)$. However, in $T(VII_h)$ ${\bf N}_{\times}=0$ may occur.
If one wants to find the equilibrium points and investigate their
stability then this choice of gauge is particularly useful.
Equilibrium points in the neighbourhood of ${\bf N}_{\times}=0$
have to be treated separately.

\subsubsection{``${\mbold\Sigma}_1/{\bf v}$-gauge'': ${\mbold\Sigma}_1/{\bf v}$ purely real} These choices of gauge are related and use the function $\phi$ to simplify the function ${\mbold\Sigma}_1$ or ${\bf v}$. For example, we can choose ${\rm Im}({\mbold\Sigma}_1)=0$ or ${\rm Im}({\bf v})=0$. Both of these choices  can be achieved by choosing $\phi$ to obey
\beq \phi'=\sqrt{3}\left[\Sigma_{23}+(\bar{N}+N_-)v_1\right]. \eeq
The remaining constant $\phi$ transformation can then be used to
choose either ${\mbold\Sigma}_1$ or ${\bf v}$ purely real
(however, in general  both of them cannot be purely real). For
this gauge, all of the variables are bounded and well-behaved.
However, for $T_1(\mathcal{A})$ the gauge choice becomes
degenerate so that there is still an undetermined gauge for this
invariant subspace. We observe that this invariant subspace
contains all of the non-tilted  and the vacuum models; hence, for
equilibrium points on the vacuum boundary, or non-tilted ones,
this choice of gauge is not very well suited. However, in
$T(\mathcal{A})- T_1(\mathcal{A})$, this choice of gauge completely
fixes the gauge and is well-behaved; this choice was used in the tilted type II analysis \cite{HBWII}.

\subsection{Equilibrium points in class B spacetimes}
A number of important equilibrium point are given in the text
below. Equilibrium points in the invariant Bianchi type VI$_h$ 
have been studied in ref. \cite{Apo}. This case contains many
equilibrium points and we shall only present the plane-wave equilibrium points explicitly here (see section \ref{pwaveVI}).

The following result is useful in the analysis of the evolution in the invariant type VII$_h$ set, $T(VII_h)$, and the type V set, $T(V)$:
\begin{thm}[Type V and VII$_h$ equilibrium points]
All equilibrium points of type V (${\bf N}_{\times}=\bar{N}=0$,
$A\neq 0$) and VII$_h$ ($\left|{\bf
N}_{\times}\right|^2-\bar{N}^2<0$, $A\neq 0$) have
${\mbold\Sigma}_1={\bf v}=0$; i.e., all equilibrium points of type
V and VII$_h$ lie in the invariant subspaces $T_1(V)$ and
$T_1(VII_h)$, respectively. Furthermore, equilibrium points in the
interior of $T_1(VII_h)$ are either plane-wave spacetimes or FRW
universes. \label{thm:V+VIIh}
\end{thm}
\begin{proof}
The type V and VII$_h$ cases need to be treated separately.

\textit{Type V}: Assuming ${\bf v}\neq 0$ leads, after a fairly
straightforward calculation, to a contradiction. Thus we must have
${\mbold\Sigma}_1={\bf v}=0$.

\textit{Type VII$_h$}: The proof is somewhat lengthy and an outline of the proof is given in  Appendix \ref{App:Proof}.
\end{proof}

For $\gamma>2/3$ the (non-vacuum) FRW universes are unstable. Therefore we immediately have:
\begin{cor}
In the case of Bianchi type VII$_h$ models the only future attracting equilibrium points for non-inflationary fluids ($\gamma>2/3$) are the plane-wave solutions.
\end{cor}
We also note that for inflationary fluids, $0<\gamma<2/3$, the FRW
universe ($\Omega=1$, $V=0$) is stable in $T(\mathcal{A})$. In fact, for $0<\gamma\leq 6/7$ (!!) there exists a monotonically increasing function $Z_1$, defined by
\beq
Z_1&\equiv& \alpha\Omega^{1-\gamma}, \quad \alpha=\frac{(1-V^2)^{\frac 12(2-\gamma)}}{G^{1-\gamma}_+V^{\gamma}_{\phantom{+}}}, \\
Z_1'&=&\left[2(1-\gamma)q+(2-\gamma)+{\gamma}\mathcal{S}\right]Z_1.\nonumber 
\eeq
To see that this is a monotonically increasing function, note first that $|\mathcal{S}|\leq 2\Sigma$ \cite{hervik}. Then, using the constraint equation (\ref{const:H}) and $(1-\Sigma)\geq (1-\Sigma^2)/2$, we can write
\beq
&& 2(1-\gamma)q+(2-\gamma)+{\gamma}\mathcal{S}\nonumber\\ 
&&\geq (6-7\gamma)\Sigma^2+2(1-\gamma)(\left|{\bf N}_{\times}\right|^2+A^2)+\frac{\gamma(1-\gamma)(V^2+3)}{G_+}\Omega , 
\eeq
which is strictly positive for $\gamma<6/7$. 
Thus for $0<\gamma\leq 6/7$  $Z_1$ is monotonically increasing as claimed. 
\begin{thm}\label{thm:nontilted}
For $0<\gamma\leq 6/7$, all tilted Bianchi models  (with $\Omega>0$, $V<1$) of solvable type are asymptotically non-tilted at late times. 
\end{thm}
\begin{proof}
Use of the monotonic function $Z_1$. 
\end{proof}
In fact, the result that these models are asymptotically non-tilted at late times is true for all $\gamma<1$; this follows from an analysis, but is not covered by this theorem. This theorem is, in fact, true for all ever-expanding Bianchi  models, including ever-expanding class A models\footnote{For class A models ($A=0$) the upper bound $6/7$ can be slightly improved to $14/15$ by modifying the monotonic function $Z_1$ to 
\[\tilde{Z}_1\equiv \tilde{\alpha}\Omega^{1-\beta}, \quad \tilde{\alpha}=\frac{(1-V^2)^{\frac 12(2-\gamma)}}{G^{1-\beta}_+V^{\beta}_{\phantom{+}}}, \quad \beta=\frac 67\gamma .\] }. 

An immediate result of the above is the following corollary:
\begin{cor}[Cosmic no-hair]
For $\Omega>0$, $V<1$, and $0<\gamma<2/3$ we have that
\[ \lim_{\tau\rightarrow \infty}\Omega=1, \quad \lim_{\tau\rightarrow \infty}V=0.\]
\label{no-hair}
\end{cor}
\begin{proof}
Using Theorem \ref{thm:nontilted} we can write 
\[ \Omega'=\Omega\left[2q-(3\gamma-2)+\delta\right] \]
 where $\delta\rightarrow 0$ as $\tau\rightarrow\infty$. Furthermore, for $0<\gamma<2/3$,
\[ 2q-(3\gamma-2)\geq(2-3\gamma)(1-\Omega)\geq 0, \]
where equality can only occur for $\Omega=1$. Assuming $\Omega<1$, then for every $0<\gamma<2/3$ there will exist a $\tau_1$ such that $2q-(3\gamma-2)+\delta >0$ for all $\tau>\tau_1$. The corollary now follows. 
\end{proof}
This result is usually referred to as the cosmic no-hair theorem; the original version given by Wald \cite{Wald} was for a cosmological constant. The above corollary generalises the non-tilted result given in \cite{DS1}. We shall be primarily interested in the future asymptotic behaviour of models with $2/3< \gamma<2$ henceforward. 
 
\section{Special subsets}

\subsection{Bianchi Type V}
The simplest model of class B is the type V model for which ${\bf
N}_{\times}=\bar{N}=0$, $A\neq 0$. Hewitt and Wainwright
\cite{HWV} studied a subclass of this model, namely the evolution
in the $T_1(V)$ invariant subspace.

For the type V models the constraint (\ref{const:v2}) simplifies to
\beq
\sqrt{3}A{\mbold\Sigma}_1=\frac{\gamma\Omega}{G_+}{\bf v},
\eeq
which implies that ${\mbold\Sigma}_1$ and ${\bf v}$ are parallel. Thus for type V the ${\mbold\Sigma}_1$ and ${\bf v}$ gauges are equivalent.

Theorem \ref{thm:V+VIIh} implies that all of the equilibrium
points of type V are given in Hewitt and Wainwright's analysis
\cite{HWV}. Hence, the equilibrium points can be extracted
directly from that paper. The stability analysis, on the other
hand, must be performed in the full state space which will change
the stability properties of the equilibrium points.

The equilibrium points important for the late-time behaviour in
the invariant space $T_1(V)$ are as follows:
\begin{enumerate}
\item{} ${\mathcal M}$, Milne: Future stable in $T_1(V)$ for $2/3<\gamma<4/3$.
\item{} $\widetilde{\mathcal M}$, ``Tilted Milne'': Future stable in $T_1(V)$ for $4/3<\gamma<2$.
\item{} ${\mathcal M}_{\pm}$, ``Extremely tilted Milne'': ${\mathcal M}_{-}$ is future stable in $T_1(V)$ for $6/5<\gamma<2$.  ${\mathcal M}_{+}$ unstable for $0<\gamma<2$.
\end{enumerate}

By considering these equilibrium points in the full type $T(V)$ state space we obtain the following stability properties:
\begin{enumerate}
\item{} ${\mathcal M}$: Future stable in $T(V)$ for $2/3<\gamma<4/3$.
\item{} $\widetilde{\mathcal M}$: Unstable in $T(V)$.
\item{} ${\mathcal M}_{\pm}$: ${\mathcal M}_{-}$ is future stable in $T(V)$ for $6/5<\gamma<2$.  ${\mathcal M}_{+}$ unstable for $0<\gamma<2$.
\end{enumerate}

As we shall see later, these results can be directly extracted
from the plane wave analysis (even though we assume ${\bf
N}_{\times}\neq 0$ in this case). The reason for this is that the
Milne universe can be considered the isotropic limit of the vacuum
plane waves ($\Sigma_+, \bar{N}\rightarrow 0$).

\subsection{Irrotational Bianchi type VII$_0$}
Let us take a first look at the tilted Bianchi type
VII$_0$ model. For the non-tilted model it is known that the
variable $\bar{N}$ diverges as $\tau\rightarrow \infty$
\cite{VII0,VII0rad} which causes the non-tilted type VII$_0$ models to experience a \emph{self-similarity breaking} at late times \cite{VII0,VII0rad}. We will show that this self-similarity breaking persists within the subspace $T_1(VII_0)$. 
 In
the non-tilted case, the variable ${\bf N}_{\times}$ will
oscillate around the value zero. To avoid the problem of the ${\bf
N}$-gauge at  ${\bf N}_{\times}=0$, we will choose the F-gauge
for which $\phi'=0$.

We will consider the invariant subspace $T_1(VII_0)$, where
${\bf v}={\mbold\Sigma}_1=0$, and we will assume, with no loss of generality,   that $\bar{N},v_1>0$. In this invariant subspace there are
two  monotonic functions, 
\beq Z_2\equiv\frac{|{\bf
N}_{\times}|^2+|{\mbold\Sigma}_{\times}|^2}{\bar{N}^2+|{\mbold\Sigma}_{\times}|^2},&&  \frac{Z_2'}{Z_2}=-\frac{4|{\mbold\Sigma}_{\times}|^2\left(\bar{N}^2-|{\bf
N}_{\times}|^2\right)(\Sigma_++1)}{\left(|{\bf
N}_{\times}|^2+|{\mbold\Sigma}_{\times}|^2\right)\left(\bar{N}^2+|{\mbold\Sigma}_{\times}|^2\right)}, \nonumber \\
Z_3\equiv \frac{\left(\bar{N}^2-|{\bf N}_{\times}|^2\right)^{n}\beta\Omega}{\left(1+n\Sigma_+\right)^{2(1+n)}},&&  \frac{Z_3'}{Z_3}=\frac{4\left[\left(\Sigma_++n\right)^2+(1-n^2)\left(|{\mbold\Sigma}_{\times}|^2+\frac{(1+2n)\Omega V^2}{3G_+}\right)\right]}{1+n\Sigma_+},\nonumber  \\ &&
\eeq 
where 
\beq \beta\equiv\frac{(1-V^2)^{\frac12(2-\gamma)}}{G_+}, \quad n\equiv\frac 14(3\gamma-2).\label{def:beta}\eeq
\begin{thm}
For a non-inflationary perfect fluid ($2/3<\gamma<2$) and any initial value in $T_1(VII_0)$ with $V<1$ and $\Omega>0$, we have that
\[ \lim_{\tau\rightarrow+\infty}|\bar{N}|=\infty.\]
\end{thm} 
\begin{proof} 
Use of the monotonic function $Z_3$ (the proof is completely analogous to the non-tilted case \cite{VII0}). 
\end{proof}
We note that this implies that these models are not asymptotically self-similar at late-times. Let us thus assume
that $1/\bar{N}\rightarrow 0$ and investigate what consequences
this may have for the evolution of the tilted type VII$_0$ model.
For $\bar{N}\rightarrow \infty$, the equations for ${\bf
N}_{\times}$ and ${\mbold\Sigma}_{\times}$ have the asymptotic form \beq
{\bf N}_{\times}'&=&
2\sqrt{3}\bar{N}{\mbold\Sigma}_{\times}, \nonumber \\
{\mbold\Sigma}_{\times}'&=& -2\sqrt{3}\bar{N}{\bf N}_{\times}.
 \eeq
These equations can be solved in quadrature. By defining
$\varphi\equiv \int_{\tau_0}^{\tau}2\sqrt{3}\bar{N}d\tau$, we have
the asymptotic form \beq {\bf N}_{\times}=\frac
12(\alpha_0e^{-i\varphi}+\beta_0e^{i\varphi}), \qquad
{\mbold\Sigma}_{\times}=\frac
1{2i}(\alpha_0e^{-i\varphi}-\beta_0e^{i\varphi}). \eeq We also
note that if we require ${\bf N}_{\times}\neq 0$ for all $\tau$,
then $|\alpha_0|\neq |\beta_0|$.

We have a similar type of oscillation in this case. This can be
verified by calculating the scalar $|{\mbold\Sigma}_{\times}|^2$
: \beq |{\mbold\Sigma}_{\times}|^2=\frac
14\left(|\alpha_0|^2+|\beta_0|^2-2|\alpha_0||\beta_0|\cos
2\varphi\right). \eeq
We also note that 
\beq \mathrm{Im}({\mbold\Sigma}_{\times}{\bf
N}_{\times}^*)=\frac 14\left(|\alpha_0|^2-|\beta_0|^2\right), 
\eeq
so that according to the constraint (\ref{const:v1}), the type
VII$_0$ model ($A=0$) has non-zero $v_1$ if $|\alpha_0|\neq
|\beta_0|$ in the limit $\bar{N}\rightarrow\infty$.

Based on these initial investigations, we introduce the four scalars 
\beq
\sigma_1=\left|{\bf N}_{\times}\right|^2+\left|{\mbold\Sigma}_{\times}\right|^2,&& \quad \sigma_2=2\mathrm{Im}({\mbold\Sigma}_{\times}{\bf
N}_{\times}^*) \nonumber \\
\sigma_3=\left|{\bf N}_{\times}\right|^2-\left|{\mbold\Sigma}_{\times}\right|^2,&& \quad \sigma_4=2\mathrm{Re}({\mbold\Sigma}_{\times}{\bf
N}_{\times}^*).
\eeq
At late times, the scalars $\sigma_3$ and $\sigma_4$ will oscillate with increasing frequency. Following \cite{VII0}, we incorporate this oscillation into a phase $\psi$ by defining
\beq
\sigma_3=\rho\sin\psi, \quad \sigma_4=\rho\cos\psi.
\eeq
Furthermore, by defining $M\equiv 1/\bar{N}$, the equations of motion can be written
\beq
\Sigma_+'&=& (Q-2)\Sigma_+-\sigma_1-\frac{\gamma\Omega V^2}{G_+}-(1+\Sigma_+)\rho\sin\psi,\label{eq:VII0Sigma} \\
\sigma_1'&=& 2(Q+\Sigma_+-1)\sigma_1+\left(\Sigma_++1-\sigma_1\right)\rho\sin\psi, \\
V'&=& \frac{V(1-V^2)}{1-(\gamma-1)V^2}\left[(3\gamma-4)+2\Sigma_+\right], \\
\rho'&=& 2\left[(Q+\Sigma_+-1)\rho+(\Sigma_++1)\sigma_1\sin\psi-\rho^2\sin\psi\right], \\
M'&=& -M\left(Q+2\Sigma_+-\rho\sin\psi+\sqrt{3}M\rho\cos\psi\right), \\
\psi'& =& \frac{1}{M}\left[4\sqrt{3}+2M(\Sigma_++1)\frac{\sigma_1}{\rho}\cos\psi\right],
\label{eq:VII0M}\eeq
where 
\beq
Q=2\Sigma_+^2+\sigma_1+\frac 12\frac{(3\gamma-2)+(2-\gamma)V^2}{G_+}\Omega.
\eeq
The variables $\Omega$ and $\sigma_2$ are determined from the constraints, 
\beq
\Omega=1-\sigma_1-\Sigma_+^2, \quad \sigma_2=-\frac{\gamma\Omega V}{G_+}.
\eeq
Moreover, the identity (\ref{eq:NSconstraint}) implies the constraint
\beq 
\sigma_1^2=\sigma_2^2+\rho^2. 
\label{const:VII0id}
\eeq
The analysis in \cite{VII0} showed that for the non-tilted model, effectively the oscillatory terms in the  system of equations are dynamically negligible at late times. A similar analysis  shows that the same is true in this case. This means that at late times we obtain a reduced system of equations by simply dropping the terms containing $\sin\psi$ and $\cos\psi$. This enables us to state:
\begin{thm}[Global attractor for $2/3<\gamma<4/3$]
For  $2/3<\gamma<4/3$, a non-LRS universe in $T_1(VII_0)$ with $V<1$ and $\Omega>0$ has
\beq
\lim_{\tau\rightarrow\infty}(\Sigma_+,\sigma_1,\rho,V)=(0,0,0,0), \quad \lim_{\tau\rightarrow\infty}\Omega=1.
\eeq
\end{thm}
\begin{proof}
The proof is completely analogous to that in \cite{VII0} (by replacing $\Omega$ with $\beta\Omega$, where $\beta$ is defined in eq.(\ref{def:beta})), or using $\hat{Z}_4$ defined in Appendix \ref{proof:VII0}.
\end{proof}

When the fluid is stiffer than radiation ($4/3<\gamma<2$), the universe will not isotropise in terms of the shear. By studying the reduced system of equations -- which is obtained by simply dropping the terms containing sine and cosine in eqs.(\ref{eq:VII0Sigma})--(\ref{eq:VII0M}) -- we can show that a solution asymptotes to a point on the following line-bifurcation:
\beq&&
\Sigma_+\rightarrow -\frac 12(3\gamma-4), \quad Q\rightarrow \frac 12(3\gamma-2), \nonumber \\
&& \sigma_1\rightarrow  \frac{3(2-\gamma)}{4(1-V^2)}\left[(3\gamma-4)-V^2(5\gamma-4)\right],\quad \sigma_2\rightarrow  -\frac{3V\gamma(2-\gamma)}{2(1-V^2)}, \nonumber \\
&& \rho \rightarrow  \frac{3(2-\gamma)}{4(1-V^2)^{\frac 12}}\sqrt{(3\gamma-4)^2-V^2(5\gamma-4)^2},\quad \Omega\rightarrow \frac{3(2-\gamma)}{2(1-V^2)}\left[1+(\gamma-1)V^2\right], \nonumber \\
&& \qquad\quad  0\leq V\leq \frac{3\gamma-4}{5\gamma-4}.
\label{eq:VII0line}\eeq
\begin{thm}[Global attractors for $4/3<\gamma<2$] \label{thm:VII0>rad}
For  $4/3<\gamma<2$, a non-LRS universe in $T_1(VII_0)$ with $V<1$ and $\Omega>0$ has
\beq
\lim_{\tau\rightarrow\infty}\Sigma_+=-\frac 12(3\gamma-4),
\eeq
and consequently approaches  the line given by eq.(\ref{eq:VII0line}). 
\end{thm}
\begin{proof}
A proof is provided in Appendix \ref{proof:VII0}.
\end{proof}

To summarise, these results imply that for perfect fluids with $2/3<\gamma<4/3$ the $T_1(VII_0)$ models isotropise in terms of the shear, while for fluids with $4/3<\gamma<2$, the $T_1(VII_0)$ models \emph{do not} isotropise in terms of the shear. This result is similar to the non-tilted case \cite{VII0}. Regarding the expansion-normalised  Weyl tensor we thus expect a similar behaviour as in the non-tilted case. In the terminology of \cite{BHWeyl} the $T_1(VII_0)$ models are, for fluids stiffer than dust ($\gamma=1$), \emph{extremely Weyl dominant} at late times. Note that the radiation case, $\gamma=4/3$, has not been treated here; this case requires a separate analysis \cite{VII0rad}. 

In the full tilted Bianchi VII$_0$ case we expect $\bar{N}\rightarrow\infty$ and a similar behaviour to occur. Tilted type VII$_0$ models is presently being studied by other authors \cite{WCL}.

\section{Stability of the plane waves}
In this section we will address the question: Are the vacuum plane
waves stable in the set of tilted Bianchi models?  To answer this
question we must first define what we mean by the ``vacuum plane
waves''. These plane-wave solutions correspond to an invariant set, $\mathcal{P}(\mathcal{A})$, of the dynamical system. Since this set
is vacuum, $\Omega=0$, the tilt velocity decouples from the
equations of motion and becomes unphysical in the invariant set itself. However, a perturbation away from the vacuum
solutions  generically has $\Omega\neq 0$, and in this case it is
meaningful to talk about tilt. Hence, we may have a situation
where the asymptotic solutions are identical but where the
asymptotic tilt velocities are different.

To see how the system of equations decouples near vacuum
equilibrium points,  we linearise the full system of equations
with respect to an equilibrium point with $\Omega=0$. Let ${\sf
Y}$ be a column vector corresponding to the variables given by the
set of eqs. (\ref{eq:Sigma+eq})-(\ref{eq:Aeq}), and ${\sf v}$ the
vector $(v_1,v_2,v_3)$. Furthermore, we assume that the constraint
equations have been used to reduce the vector ${\sf Y}$. Then, at
the linearised level, we can write \beq
\begin{bmatrix}
\delta {\sf Y} \\ \delta\Omega \\ \delta {\sf v}
\end{bmatrix}' =
\begin{bmatrix}
 {\sf A} & \lambda_{{\sf Y}} & 0 \\
0 & \lambda_{\Omega} & 0 \\
{\sf C} & \lambda_{{\sf v}} & {\sf B}
\end{bmatrix}
\begin{bmatrix}
\delta {\sf Y} \\ \delta\Omega \\ \delta {\sf v}
\end{bmatrix}.
\eeq We note that the eigenvalue equation reduces to 
\beq
\det\left({\sf A}-\Lambda{\sf 1}_1\right)\det\left({\sf
B}-\Lambda{\sf 1}_2\right)(\lambda_{\Omega}-\Lambda)=0. 
\eeq
Hence, near the vacuum equilibrium points, the system of equations
decouples and we can study the two decoupled systems separately. In
addition, we need to check the value of $\lambda_{\Omega}$ to make
sure that the vacuum approximation remains valid.

Since we are dealing with an invariant set for which $\Omega=0$,
it is convenient to choose the ${\bf N}_{\times}$-gauge, where the
case $\lambda =\infty$ correspond to $N_{23}=0$. We recall that in
general $\bar{N}=\lambda N_{23}$.

By vacuum plane waves we will mean all solutions where:
\paragraph{{\bf Vacuum plane waves:}} $\mathcal{P}(\mathcal{A})$, ($0<\gamma<2$)
\beq q=-2\Sigma_+, \quad N_{23}^2=-\Sigma_+(1+\Sigma_+), \quad A^2=(1+\Sigma_+)^2,
\nonumber \\ \Sigma_-^2=-\Sigma_+(1+\Sigma_+), \quad \Sigma_{12}=\Sigma_{13}=\Sigma_{23}=\Omega=0.
\eeq
The parameter $h$ is determined by $\lambda $ via
\[ 3h\Sigma_+(1-\lambda^2)=(1+\Sigma_+),\]
and  $\Sigma_+$ is bounded by
\beq
-1< \Sigma_+ \begin{cases}
<0,& \quad  \lambda^2>1~ (\text{VII}_h),\\
<0, & \quad \lambda^2=1~(\text{IV}),\\
\leq \frac{1}{3h-1},&\quad  \lambda^2<1~(\text{VI}_h).
\end{cases}
\eeq Without loss of generality we can also  assume that
$A,\Sigma_-,N>0$.

To investigate the various versions of the vacuum plane waves
(i.e., with different tilt velocities) we can assume that the
geometric variables have the above values and use these in the
equations for $v_1$, $v_2$, $v_3$. The paradigm is that as the universe approaches a plane wave, the variables $\Sigma_+$, $\Sigma_-$, etc. 'freeze in' and effectively become constants. There are several results that support this paradigm. A stability analysis shows that this is true locally. Furthermore, comprehensive numerical analysis in the full state space suggests that this is  true globally; indeed, evidence for other behaviour has not been seen \cite{HHC}.  
Once these variables are frozen, we obtain a reduced
three-dimensional dynamical system that can be studied. These
equations can therefore be reinterpreted as describing the tilt
velocity of test matter in a plane-wave background. For
convenience, and to distinguish them from the full set of
equations, we define $(x,y,z)\equiv
\left(v_1,\frac{v_2+v_3}{\sqrt{2}},\frac{v_2-v_3}{\sqrt{2}}\right)$.
The reduced equations for $(x,y,z)$ are then: \beq
x'&=& (T+2\Sigma_+)x-A(y^2+z^2)-2\sqrt{3}N_{23}yz, \nonumber \\
y'&=& (T-\Sigma_++Ax)y-\alpha(1-\lambda)z, \nonumber \\
z'&=& (T-\Sigma_++Ax)z-\alpha(1+\lambda)y,
\label{eq:vSystem}\eeq
where $\alpha=\sqrt{3}(\Sigma_--xN_{23})=(1-x)\sqrt{3}\Sigma_-$, since $N_{23}=\Sigma_-$ for the plane-wave solutions. The function $T$ is given by eq.(\ref{eq:defs}) with
\[ V^2=x^2+y^2+z^2, \quad V^2\mathcal{S}=\Sigma_+(-2x^2+y^2+z^2)+2\sqrt{3}\Sigma_-yz.\]
The requirement $0\leq V\leq 1$, implies
\[ 0\leq x^2+y^2+z^2 \leq 1. \]
Hence, the state space is the solid unit ball, $D^3$.

To get some sense  of the resulting dynamics, we note that 
\beq
R'=(T-\Sigma_++Ax)R, \qquad
R^2\equiv(\lambda+1)y^2+(\lambda-1)z^2. \eeq 
Hence, $R,x=$constant
describes ellipses for type VII$_h$ ($\lambda^2>1$), straight
lines for type IV ($\lambda^2=1$), and hyperbolae for type VI$_h$
($\lambda^2<1)$. (Note that since these equations are not $U(1)$
invariant, there is no question that the resulting dynamics is a
gauge effect.)

Let us study the reduced dynamical system (\ref{eq:vSystem}). The
key observation is that from this reduced system, and from equation
eq.(\ref{eq:Omega}) for $\Omega$, we can say something about the
stability of the plane waves. The strategy is as follows: We first
find the asymptotic behaviour of the system (\ref{eq:vSystem}).
If, in addition, the variable $\Omega$ is stable with respect to
this asymptotic behaviour, then the vacuum plane waves will be
stable in the full set of equations. Here, we will present heuristic arguments for some of the features of these model. In a companion paper \cite{HHC} these features are more thoroughly investigated. 

The primary aim is the investigation of the type VII$_h$ plane
waves; however, it is very illustrative  (and also convenient) to
consider the type IV plane waves first. We will also consider the
type VI$_h$ plane-waves.

\subsection{Type IV plane waves}
\label{sect:IV} This case is defined by $\lambda=1$, which makes
$R$  degenerate. The invariant surface $R=y=0$, splits the solid
ball, $D^3$, into two equal halves which are related via the
symmetry $(x,y,z)\rightarrow (x,-y,-z)$. It therefore suffices to
consider the half in which  $y\geq 0$. The equilibrium points are:
\begin{enumerate}
\item{}{$\mathcal{L}(IV)$:} Non-tilted, $x=y=z=0$.
\item{}{$\tilde{\mathcal{L}}(IV)$:} Intermediately tilted, $x=\frac{3\gamma-4+2\Sigma_+}{2(\gamma-1)(1+\Sigma_+)}$, $y=z=0$, $\frac{6}{5+2\Sigma_+}<\gamma<2$.
\item{}{$\tilde{\mathcal{L}}_+(IV)$:} Extremely tilted, $x=1$, $y=z=0$.
\item{}{$\tilde{\mathcal{L}}_-(IV)$:} Extremely tilted, $x=-1$, $y=z=0$.
\item{}{$\tilde{\mathcal{F}}_{\pm}(IV)$:} Intermediately tilted, $x=-\frac{3\gamma-4-\Sigma_+}{(1+\Sigma_+)(3-2\gamma)}$,\\ $z=\pm\frac{\sqrt{(1-2\Sigma_+)(4-3\gamma)(3\gamma-4-\Sigma_+)}}{(1+\Sigma_+)(3-2\gamma)}$, $y=0$. $\frac{4+\Sigma_+}3<\gamma<\min\left(\frac 43,\frac{3}{2-\Sigma_+}\right)$.
\item{}{$\tilde{\mathcal{E}}_{\pm}(IV)$:} Extremely tilted, $x=\frac{1+\Sigma_+}{3\Sigma_+}$, $z=\pm\frac{\sqrt{(1-2\Sigma_+)(-1-4\Sigma_+)}}{3|\Sigma_+|}$, $y=0$, $-1<\Sigma_+<-\frac14.$
\end{enumerate}
For various different values of $\Sigma_+$ and $\gamma$ these
equilibrium points serve as attractors,  saddle points or
repellors. As we are mostly interested in the future asymptotic
behaviour, we will only be concerned with the future attractors. In
addition to the eigenvalues arising from the above reduced
dynamical system we have to make sure that the eigenvalue from the
$\Omega$-equation is negative as well. This will ensure that
$\Omega\rightarrow 0$ at late times so  that our approximation is
valid. Furthermore, one of the eigenvalues for
$\tilde{\mathcal{F}}_{\pm}(IV)$ and
$\tilde{\mathcal{E}}_{\pm}(IV)$ is zero (namely, the one in the
$y$-direction) so we have to go to next order to determine their
stability. This is just the reflection of the fact that
$\tilde{\mathcal{F}}_{\pm}(IV)$ and
$\tilde{\mathcal{E}}_{\pm}(IV)$ can only  be an attractor in one
of the half-spheres, and not in the other.

\begin{figure}
  \includegraphics[width=5cm]{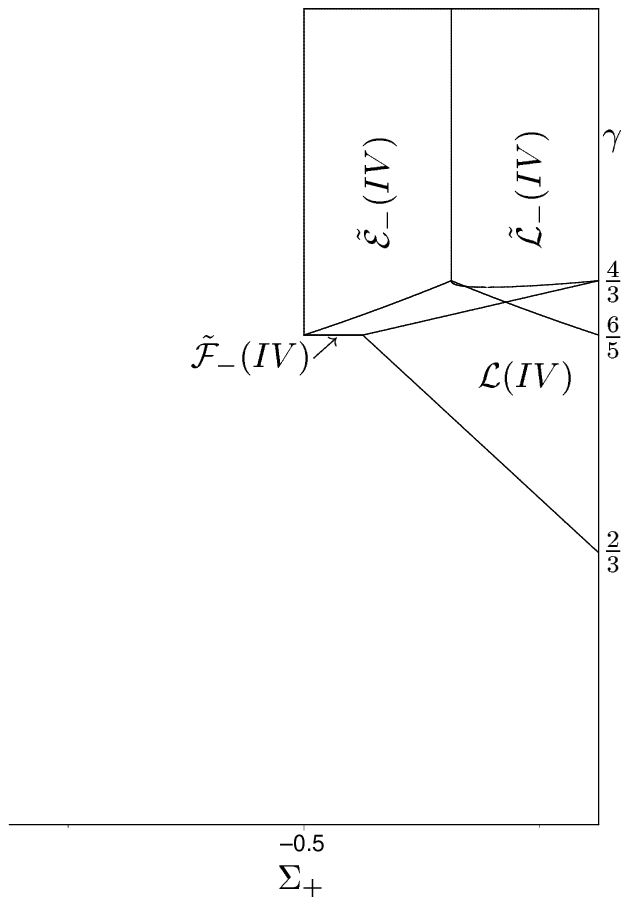}\hspace{1cm}
\includegraphics[width=5cm]{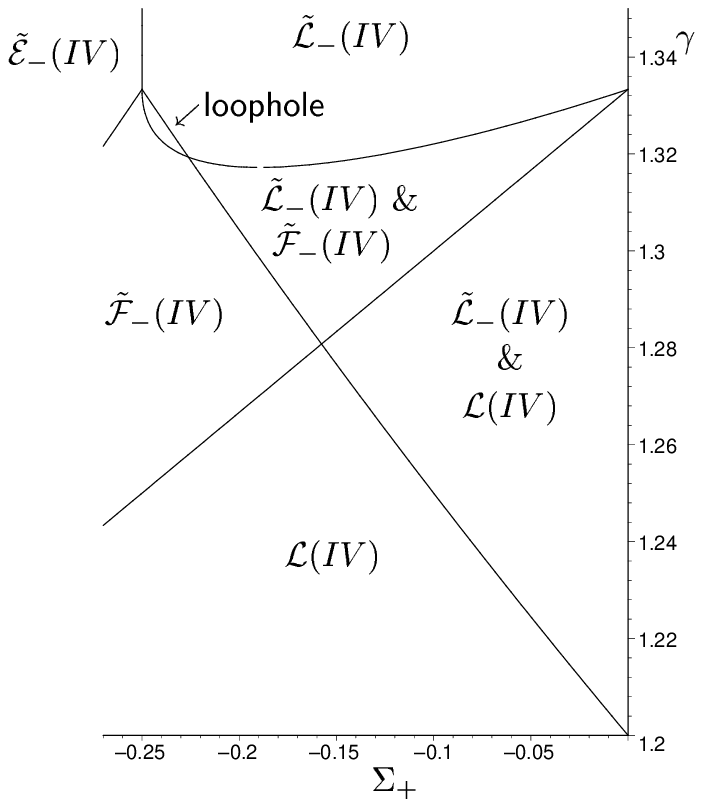}\\
  \caption{The regions where the different equilibrium points are attractors for the type IV model with $y\geq 0$.  In the left figure, all of the boundaries along the left edge mark the instability of the energy density, $\Omega$. The right figure is a magnified region of the left one showing the loophole.}\label{Fig:MapIV}
\end{figure}

\begin{figure}
  \includegraphics[width=5cm]{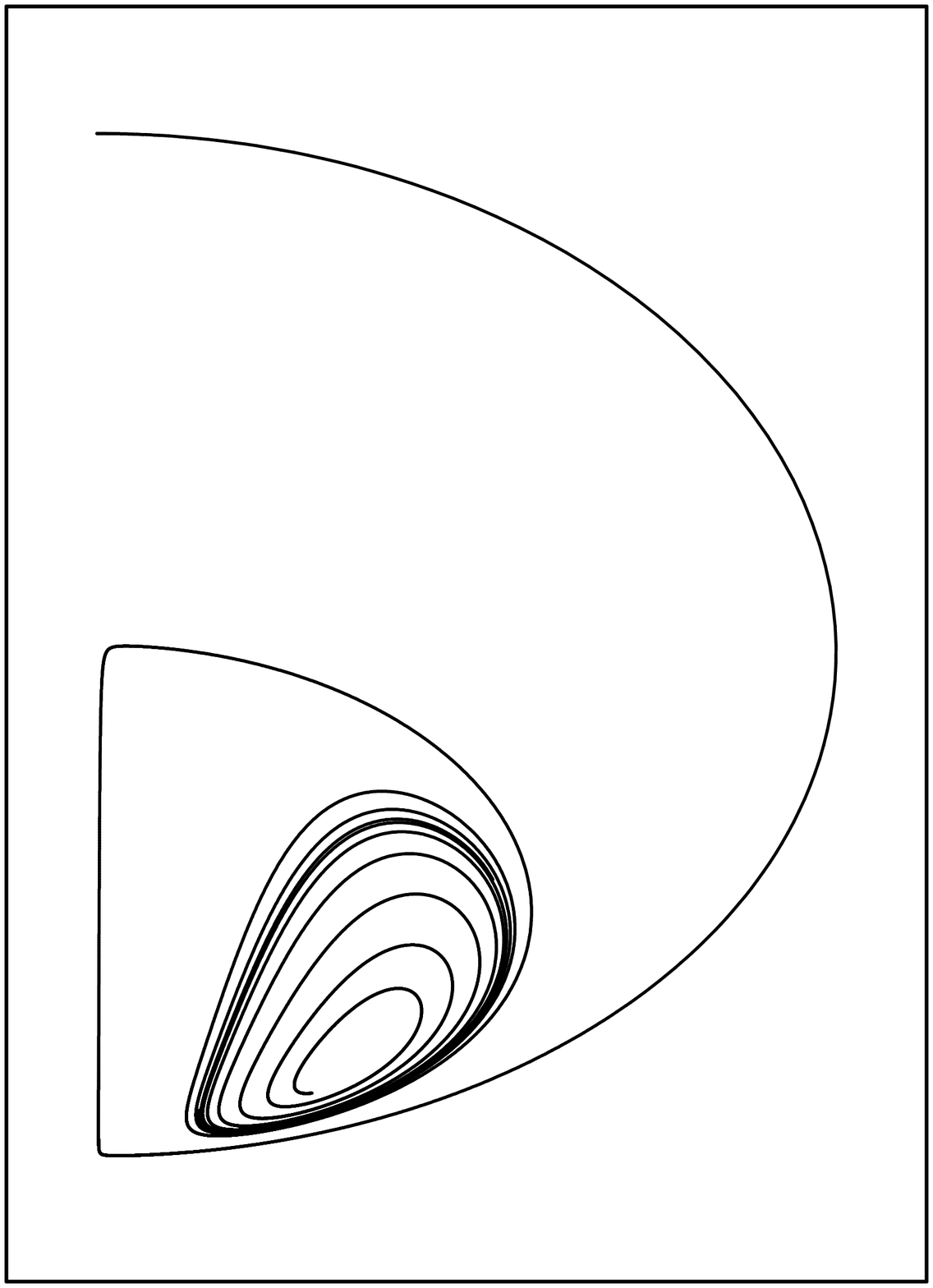}
\includegraphics[width=5cm]{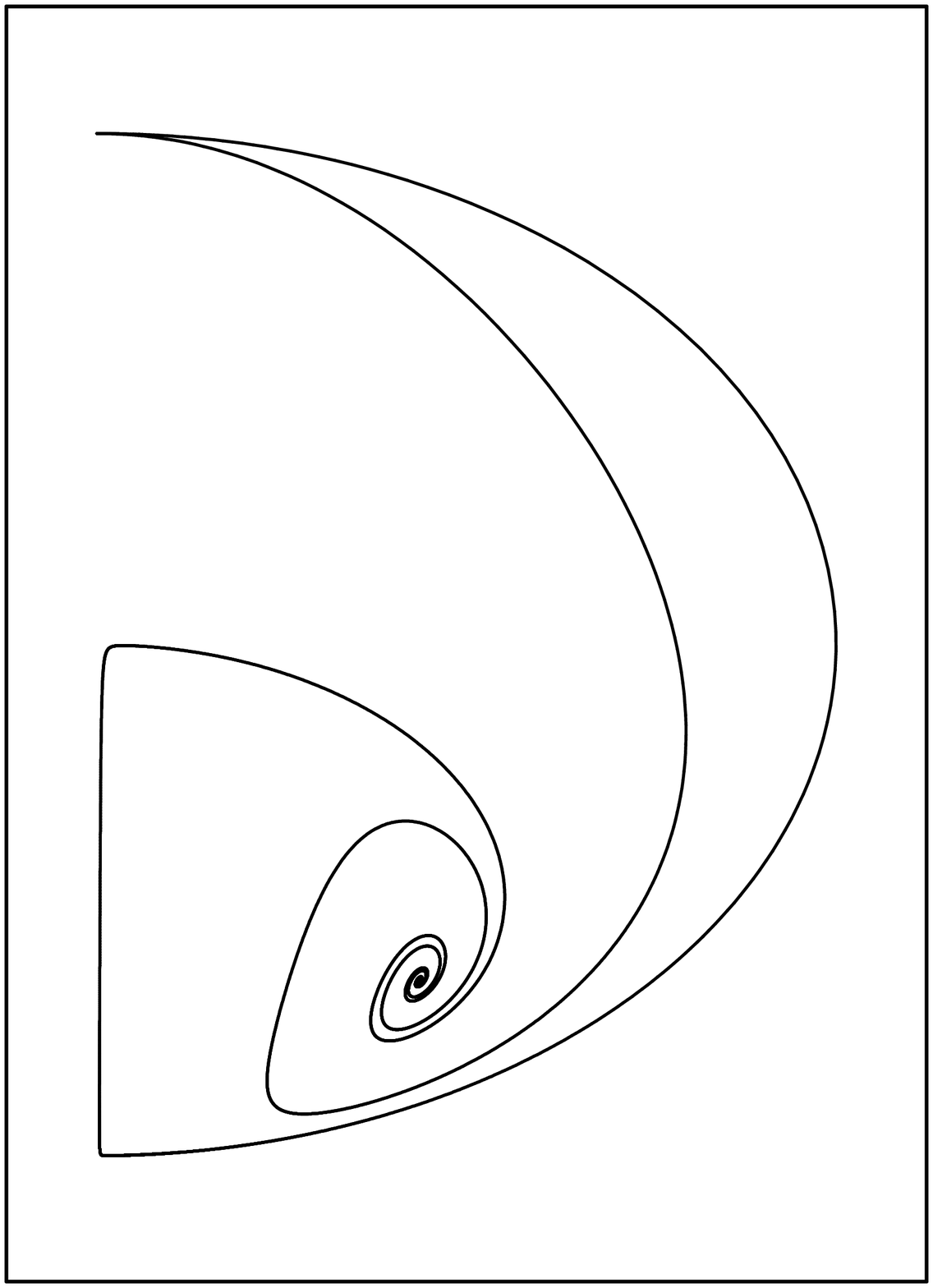}\\
  \caption{The Mussel attractor of type IV: Two solution curves where the parameters are given inside (left), and just outside (right) the loophole. The curves are projected onto the $zx$-plane. Note the  closed curve, $\mathcal{C}(IV)$, which acts as an attractor. On the right figure, the equilibrium point $\tilde{\mathcal{F}}_-(IV)$ acts as a future attractor. For both figures  $x<0$ and $z<0$ at late times.} \label{Fig:loophole}
\end{figure}

The equilibrium points are attractors for the region $y\geq 0$ for the following range of parameters:
\begin{enumerate}
\item{}{$\mathcal{L}(IV)$:} $\frac{2-4\Sigma_+}{3}<\gamma<\frac{4+\Sigma_+}3$.
\item{}{$\tilde{\mathcal{L}}(IV):$} Always unstable for $0<\gamma<2$.
\item{}{$\tilde{\mathcal{L}}_+(IV):$} Always unstable for $0<\gamma<2$.
\item{}{$\tilde{\mathcal{L}}_-(IV):$} $\frac{6}{5+2\Sigma_+}<\gamma<2$, $-\frac 14<\Sigma_+<0$.
\item{}{$\tilde{\mathcal{F}}_{\pm}(IV):$} $\tilde{\mathcal{F}}_{-}(IV)$ stable\footnote{$\tilde{\mathcal{F}}_{+}(IV)$ is stable in the same region of $\gamma$ for the half $y\leq 0$.} for $\max\left(\frac 65,\frac{4+\Sigma_+}{3}\right)<\gamma<\min\left( \gamma_0,\frac{3}{2-\Sigma_+}\right)$
\item{}{$\tilde{\mathcal{E}}_{\pm}(IV):$} $\tilde{\mathcal{E}}_{-}(IV)$ stable for  $-\frac 12<\Sigma_+<-\frac14$, $ \frac{3}{2-\Sigma_+}<\gamma<2$.
\end{enumerate}
Here, we have defined $\gamma_0$ for any given value of $\Sigma_+$
as $F(\gamma_0,\Sigma_+)=0$,  where $F(\gamma,\Sigma_+)$ is
defined in eq.(\ref{def:Fdef}) in the Appendix. Note that there is
a region where there are two co-existing future attractors; namely, the region where $\frac{6}{5+2\Sigma_+}<\gamma<\gamma_0$. Also,
there appears to be a tiny region
$\gamma_0<\gamma<\frac{6}{5+2\Sigma_+}$ (from now on called the
'loophole') which does not have any stable equilibrium points (see
Figure \ref{Fig:MapIV}).  This perhaps suggests that there is some
interesting dynamical behaviour in this region.  Numerical
experimentation seems to indicate that there is a \emph{stable
curve}, $\mathcal{C}(IV)$, which acts as an attractor in this
'loophole'. This curve lies in the invariant subspace $y=0$ and
has a 'mussel-like' shape. In Figure \ref{Fig:loophole} we have
plotted two solution curves for two different values of the
parameters $\gamma$ and $\Sigma_+$. From the figure one can
clearly see the curve, $\mathcal{C}(IV)$, which acts as a limit
cycle. Clearly it is not  sufficient  to only consider equilibrium
points as attractors when investigating the dynamics close to a
plane-wave spacetime in these SH models. These  limit cycles are dealt with more elaborately in the work \cite{HHC}.

From the above we can conclude:
For $2/3<\gamma<2$ there will always be future stable plane-wave spacetimes for the type IV model. Regarding the asymptotic tilt velocity, we  have the following possibilities depending on the value of $\gamma$.
\begin{itemize}
\item{}$2/3<\gamma<6/5$: The tilt is asymptotically zero [$\mathcal{L}(IV)$]; i.e. non-tilted.
\item{}$6/5<\gamma<4/3$: The tilt can be either zero [$\mathcal{L}(IV)$], intermediate [$\tilde{\mathcal{F}}(IV)$ or $\mathcal{C}(IV)$], or extreme [$\tilde{\mathcal{L}}_-(IV)$ or $\tilde{\mathcal{E}}(IV)$].
\item{} $4/3<\gamma<2$: The tilt is asymptotically extreme [$\tilde{\mathcal{L}}_-(IV)$ or $\tilde{\mathcal{E}}(IV)$].
\end{itemize}

\subsection{Type VII$_h$ plane waves}
In this case $\lambda^2>1$, and the level curves of $R$ describe  elliptic cylinders. We introduce elliptic polar coordinates by
\[(\sqrt{\lambda+1}y,\sqrt{\lambda-1}z)=(R\cos\varphi,R\sin\varphi),
\]
where the evolution equation for $\varphi$ is given by \beq
\varphi'=(1-x)\sqrt{3(\lambda^2-1)}\Sigma_-.
\label{eq:varphiVIIh}\eeq Since $\Sigma_-$ and $\lambda$ are
constants for the plane waves, the tilt velocity will have a
non-zero elliptic angular velocity with respect to the $x$-axis
(in agreement with theorem \ref{thm:V+VIIh}). All of the
equilibrium points must therefore lie on the $x$-axis:
\begin{enumerate}
\item{}{$\mathcal{L}(h)$:} Non-tilted, $x=y=z=0$.
\item{}{$\tilde{\mathcal{L}}(h)$:} Intermediately tilted, $x=\frac{3\gamma-4+2\Sigma_+}{2(\gamma-1)(1+\Sigma_+)}$, $y=z=0$, $\frac{6}{5+2\Sigma_+}<\gamma<2$.
\item{}{$\tilde{\mathcal{L}}_+(h)$:} Extremely tilted, $x=1$, $y=z=0$.
\item{}{$\tilde{\mathcal{L}}_-(h)$:} Extremely tilted, $x=-1$, $y=z=0$.
\end{enumerate}

One might wonder what happened to the equilibrium points
$\tilde{\mathcal{F}}_{\pm}(IV)$ and
$\tilde{\mathcal{E}}_{\pm}(IV)$ as we went over to the type
VII$_h$ models. In fact, these equilibrium points have turned into
\emph{closed periodic orbits}. Hence, also in this case, it is not sufficient to
consider only the equilibrium points; there are also exact
closed periodic orbits that may act as future attractors. These closed periodic 
orbits are difficult to deal with in full generality; however, analytical, as well as numerical results concerning the existence of these  orbits of the type VII$_h$ model are presented in \cite{HHC}.

The stability of the type VII$_h$ plane
waves appears to be similar to that of the type IV model, except that the
points $\tilde{\mathcal{F}}_{\pm}(IV)$ and
$\tilde{\mathcal{E}}_{\pm}(IV)$ have become closed orbits.
These closed orbits  appear to be stable in approximately the same
region as their IV counterparts; however, there is some
uncertainty regarding the boundary of the loophole. 

Regarding the closed curve $\mathcal{C}(IV)$ in the 'loophole',
this seems to have turned  into something more complicated for the
type VII$_h$ models. Naively, it might be anticipated that the
attracting set $\mathcal{C}(IV)$ has turned into something which
is topologically a torus, $\mathcal{T}$. Numerical work 
confirm this claim \cite{HHC}.

In any case we can conclude:
For $2/3<\gamma<2$ there will always be future stable plane-wave spacetimes for the type VII$_h$ model. Regarding the asymptotic tilt velocity, we  have the following possibilities depending on the value of $\gamma$:
\begin{itemize}
\item{}$2/3<\gamma<6/5$: The tilt is asymptotically zero [$\mathcal{L}(h)$]; i.e. non-tilted.
\item{}$6/5<\gamma<4/3$: The tilt can be either zero [$\mathcal{L}(h)$], intermediate and oscillatory, extreme  [$\tilde{\mathcal{L}}_-(h)$], or extreme and oscillatory.
\item{} $4/3<\gamma<2$: The tilt is asymptotically extreme, and can either be non-oscillatory [$\tilde{\mathcal{L}}_-(h)$] or oscillatory.
\end{itemize}

\subsection{Type VI$_h$ plane waves}\label{pwaveVI}
\begin{figure}
\includegraphics[height=5.2cm,angle=-90]{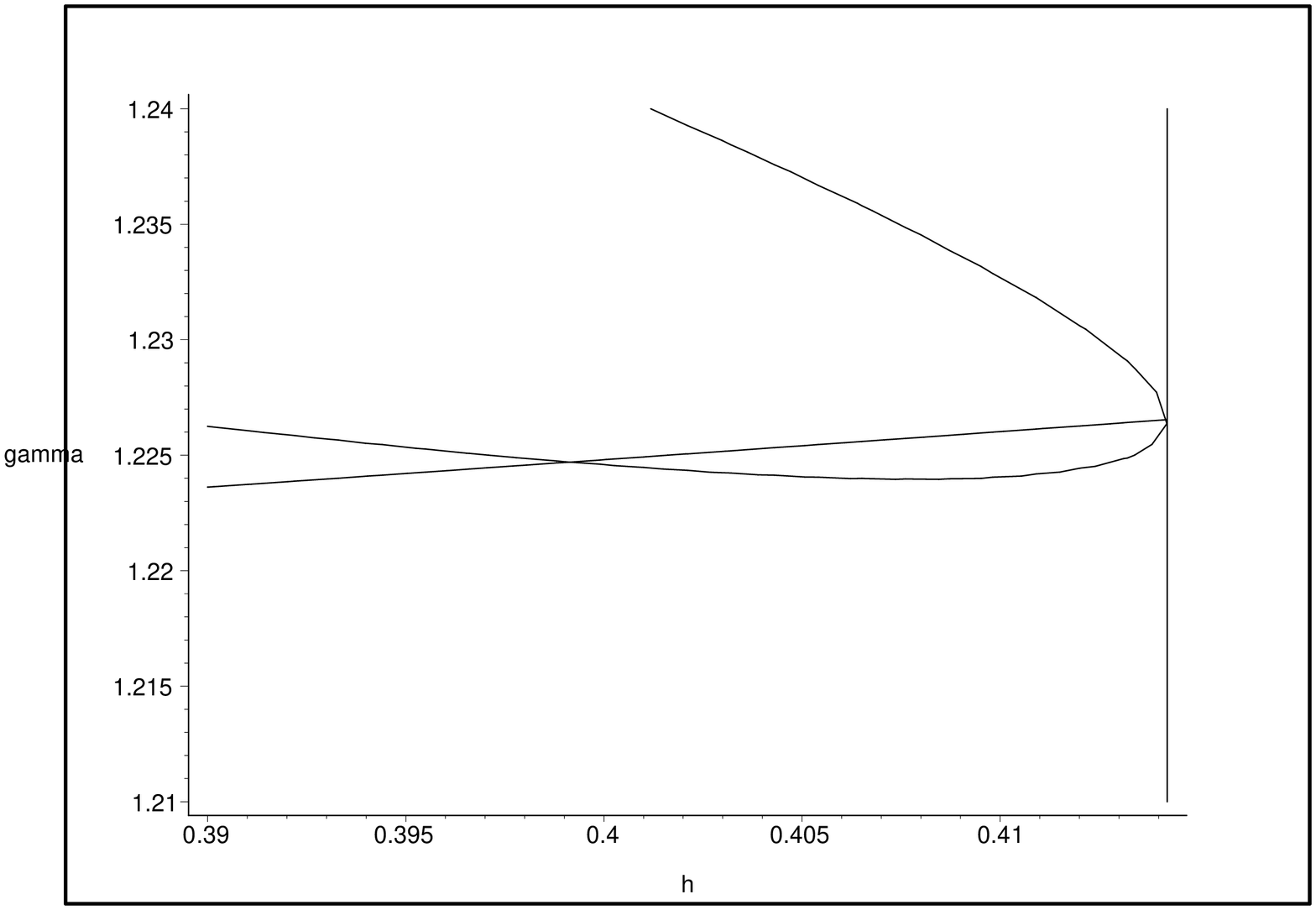}
  \includegraphics[height=5cm,angle=-90]{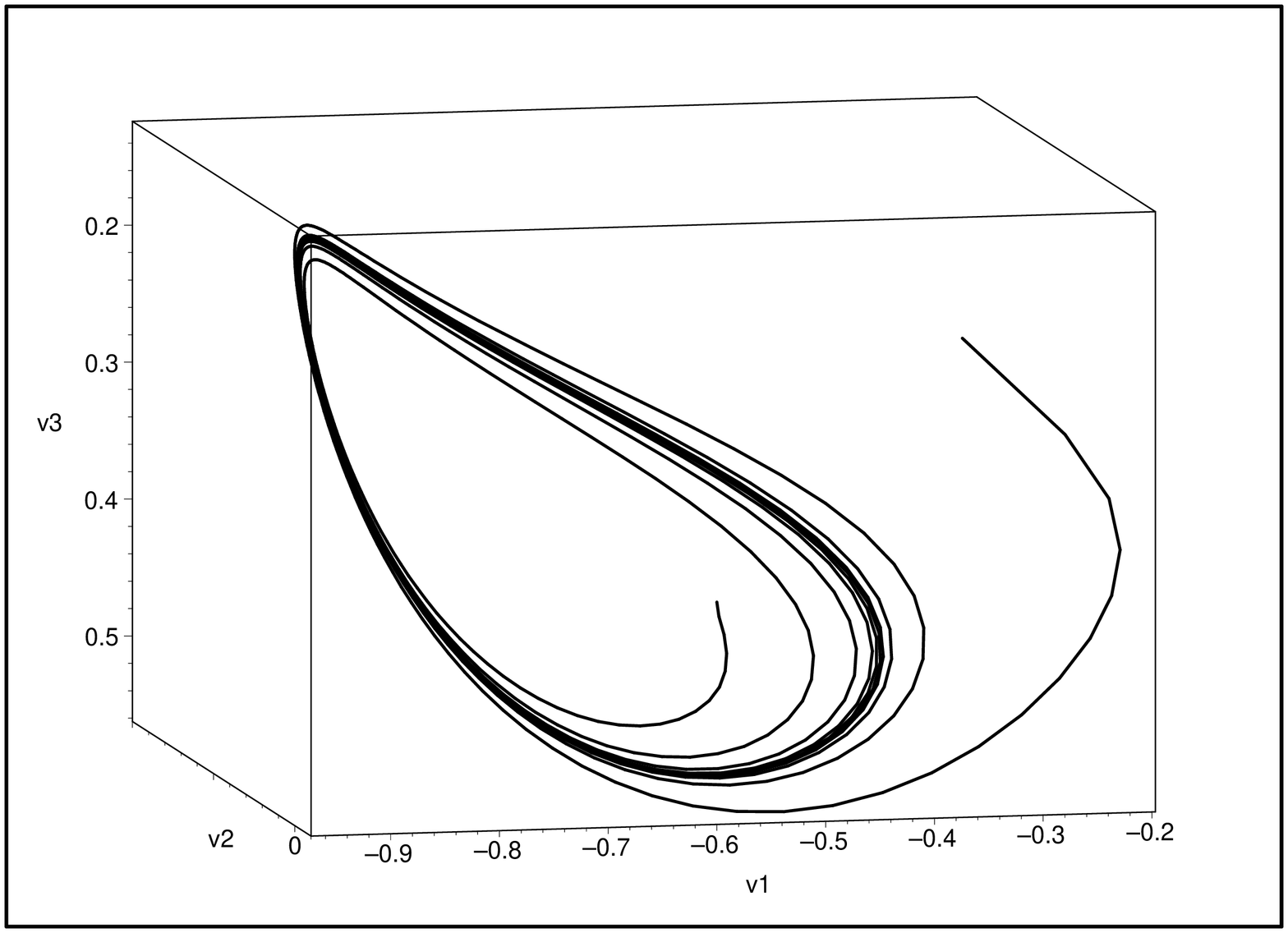}\\
  \caption{To the left a magnification of the type VI$_h$ loophole for the plane waves (where $\Sigma_+=-\tilde{h}^2/(3+\tilde{h}^2)$)
  is presented. The vertical line is $\tilde{h}=\sqrt{2}-1$, which marks the transition of stability
  between $\tilde{\mathcal{L}}_-(h)$ and $\tilde{\mathcal{E}}_-(h)$. The right picture
  shows the mussel attractor for $\tilde{h}=0.41$, $\gamma=1.225$. Two solution curves are plotted; one approaches
   $\mathcal{C}$ from the inside, while the other approaches $\mathcal{C}$ from the outside.}\label{Fig:VIhLoop}
\end{figure}

In this case we have $\lambda^2<1$, and the level curves of $R$
are hyperbolae. The invariant subspace $R=0$ defines two planes
intersecting at $y=z=0$ and divides the ball $D^3$ into 4 pieces.
Due to the symmetry $(x,y,z)\rightarrow (x,-y,-z)$, the 4
pieces are 2 equivalent pairs. Regarding the equilibrium points,
it therefore suffices to consider only equilibrium points in the
region $y\geq 0$. All of the equilibrium points reside in the
invariant subspace $R=0$ (we define $\tilde{h}=1/\sqrt{|h|}$):
\begin{enumerate}
\item{}{$\mathcal{L}(h)$:} Non-tilted, $x=y=z=0$.
\item{}{$\tilde{\mathcal{L}}(h)$:} Intermediately tilted,  $x=\frac{3\gamma-4+2\Sigma_+}{2(\gamma-1)(1+\Sigma_+)}$, $y=z=0$, $\frac{6}{5+2\Sigma_+}<\gamma<2$.
\item{}{$\tilde{\mathcal{L}}_+(h)$:} Extremely tilted, $x=1$, $y=z=0$.
\item{}{$\tilde{\mathcal{L}}_-(h)$:} Extremely tilted, $x=-1$, $y=z=0$.
\item{}{$\tilde{\mathcal{F}}_{\pm}(h):$} Intermediately tilted, 
$x=-\frac{3\gamma-4-\Sigma_+\mp(1+\Sigma_+)\tilde{h}}{(1+\Sigma_+)(3-2\gamma\pm \tilde{h})}$, $y=\chi\sqrt{1-\lambda}/\sqrt{2}$, $z=\pm\chi\sqrt{1+\lambda}/\sqrt{2}$ where \beq
\chi=\frac{\sqrt{(1-2\Sigma_+)[4-3\gamma\pm \tilde{h}(2-\gamma)][3\gamma-4-\Sigma_+\mp (1+\Sigma_+)\tilde{h}]}}{(1+\Sigma_+)(3-2\gamma\pm \tilde{h})\sqrt{1\pm\tilde{h}}}, \nonumber \\
\frac{4+\Sigma_+\pm (1+\Sigma_+)\tilde{h}}3<\gamma<\min\left(\frac {4\pm 2\tilde{h}}{3\pm \tilde{h}},\frac{3}{2-\Sigma_+\mp(1+\Sigma_+)\tilde{h}}\right).\nonumber \eeq
\item{}{$\tilde{\mathcal{E}}_{\pm}(h):$} Extremely tilted,  $x=\frac{(1+\Sigma_+)(1\pm\tilde{h})}{3\Sigma_+\pm(1+\Sigma_+)\tilde{h}}$, $y=\chi\sqrt{1-\lambda}/\sqrt{2}$, \\ $z=\pm\chi\sqrt{1+\lambda}/\sqrt{2}$ where
\beq
\chi=\frac{\sqrt{(1-2\Sigma_+)(-1-4\Sigma_+\mp2(1+\Sigma_+)\tilde{h})}}{|3\Sigma_+\pm(1+\Sigma_+)\tilde{h}|},\nonumber \\
-1<\Sigma_+<\min\left(-\frac{1\pm 2\tilde{h}}{4\pm 2\tilde{h}},~-\frac{\tilde{h}^2}{3+\tilde{h}^2}\right).\nonumber
\eeq
\end{enumerate}
Here the situation is more complicated than before.  However, from
the eigenvalues we see that the equilibrium points
$\tilde{\mathcal{F}}_{+}(h)$ and $\tilde{\mathcal{E}}_{+}(h)$
always have an unstable mode. Considering the eigenvalues for the
remaining equilibrium points we note that there are no stable
equilibrium points for $h>-1$ ($\tilde{h}>1$). For $h<-1$
($\tilde{h}<1$), there are always some stable plane-wave
equilibrium points except in a tiny region. This tiny region is
the corresponding type VI$_h$ loophole. As in the case of type IV,
in the loophole there seems to be a closed curve,
$\mathcal{C}(h)$, acting as an attractor. The curve
$\mathcal{C}(h)$ lies in the invariant subspace
$y=\chi\sqrt{1-\lambda}/\sqrt{2}$,
$z=-\chi\sqrt{1+\lambda}/\sqrt{2}$, along with all of the other
attractors. An illustration from the loophole is given in Figure
\ref{Fig:VIhLoop}.
\begin{figure}
\includegraphics[width=5.2cm]{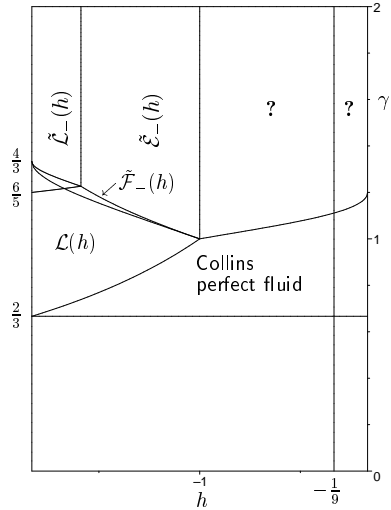}\\
  \caption{A plot of the regions of stability of the various equilibrium points.
  In the above plot we have set $\Sigma_+$ equal to the threshold value $\frac{1}{3h-1}$.
  Note that there are no stable plane-waves for $h>-1$. The loophole is approximately
  where the line $h=-3-2\sqrt{2}$ ($\tilde{h}=\sqrt{2}-1$) terminates. The question marks
  indicate regions where no stable equilibrium points are presently known.}\label{Fig:VIhMap}
\end{figure}

By a careful analysis of the remaining equilibrium points and
their eigenvalues, we obtain the following result:
\begin{itemize}
\item{} For $h<-1$ and $\gamma>\frac{2(1-h)}{1-3h}$, there are always future stable plane-wave solutions.
\item{} For $h>-1$ and $0<\gamma<2$, there are no future stable vacuum plane-wave equilibrium points.
\end{itemize}
We note that the critical unstable mode at $h=-1$ ($\tilde{h}=1$)
is the matter density $\Omega$. Furthermore, we know that there are no
other stable equilibrium points with $\Omega=0$ (for $h\neq
-1/9$). This is consequently a signal that for $h>-1$, the future
asymptote is non-vacuum.

In \cite{BHtilted} the non-tilted Collins perfect fluid solutions
of type VI$_{h}$ were investigated.  This analysis showed that for
$h>-1$, the Collins solutions were unstable against tilt whenever
$\gamma>\frac{2(3+\sqrt{-h})}{5+3\sqrt{-h}}$. This threshold value
is the line separating the regions marked ``\textsf{Collins
perfect fluid}'' and those with a question mark in Figure
\ref{Fig:VIhMap}. This instability indicates that for
$\gamma>\frac{2(3+\sqrt{-h})}{5+3\sqrt{-h}}$ the late-time
asymptote is tilted; however, no stable tilted equilibrium points are
presently known in this region.

\subsection{The exceptional case: The Collinson-French vacuum}
The plane-wave-type analysis above can also be applied to the
exceptional case of Bianchi type VI$_{-1/9}$. The non-tilted
analysis showed that the Collinson-French vacuum (or the
Robinson-Trautman solution) is stable in  the non-tilted Bianchi
cosmologies for $\gamma>10/9$ \cite{HHW}. However, the tilted
analysis showed that there is one unstable mode for $\gamma>4/3$
and one more for $\gamma>14/9$ \cite{BS,BHtilted}. Hence, a tilted
analysis is required to check the  stability.

The Collinson-French solution is given by (in the N-gauge): 
\beq
&&\Sigma_+=-\frac 13,\quad \Sigma_-=\frac{1}{3\sqrt{3}}, \quad
\Sigma_{13}=\frac{\sqrt{5}}{3\sqrt{3}},\quad
N_{23}=\frac{1}{\sqrt{2}}, \quad A=\frac{1}{\sqrt{6}},\nonumber \\
&& \Sigma_{12}=\Sigma_{23}=\Omega=\lambda=0. 
\eeq 
In this case the
$v_i$-equations also decouple and we can again treat these
separately. For the exceptional case, $h=-1/9$, which leads to the
exact vanishing of one of the constraint equations. Hence, the
tilt-velocity can only have 2 independent components and we set
$v_3=0$.

Writing $(X,Y)=(v_1,v_2)$, we obtain \beq
X'&=& \left(T-\frac 23\right)X-\frac 23\sqrt{6}Y^2, \nonumber \\
Y'&=& \left(T+\frac 23\sqrt{6}X\right)Y, \label{eq:CFequations}
\eeq where $T$ is defined as usual with
\[ V^2\mathcal{S}=\frac 23X^2. \]

The above system possesses the following equilibrium points:
\begin{enumerate}
\item{}$\mathcal{CF}$: $X=Y=0$, $0<\gamma<2$.
\item{}$\widetilde{\mathcal{CF}}_{1\pm}$: $X=-\frac{\sqrt{6}(3\gamma-4)}{2(3-\gamma)}$, $Y=\pm\frac{\sqrt{5(3\gamma-4)(3-2\gamma)}}{\sqrt{2}(3-\gamma)}$, $\frac 43<\gamma<\frac 32$.
\item{} $\widetilde{\mathcal{CF}}_{2}$: $X=\frac{\sqrt{6}(9\gamma-14)}{6(\gamma-1)}$, $Y=0$, $\frac{24-\sqrt{6}}{15}<\gamma<\frac{24+\sqrt{6}}{15}$.
\item{}$\widetilde{\mathcal{ECF}}_{\pm}$:  $X=\pm 1$, $Y=0$, $0<\gamma<2$.
\end{enumerate}

We note that  $Y=0$ is an invariant subspace of the system, which
splits the unit disc, $D^2$, into two. These two halves are
related via the symmetry transformation $Y\mapsto -Y$. Hence, it
suffices to analyse the system (\ref{eq:CFequations}) for $Y\geq
0$.

\begin{figure}
\includegraphics[width=5cm]{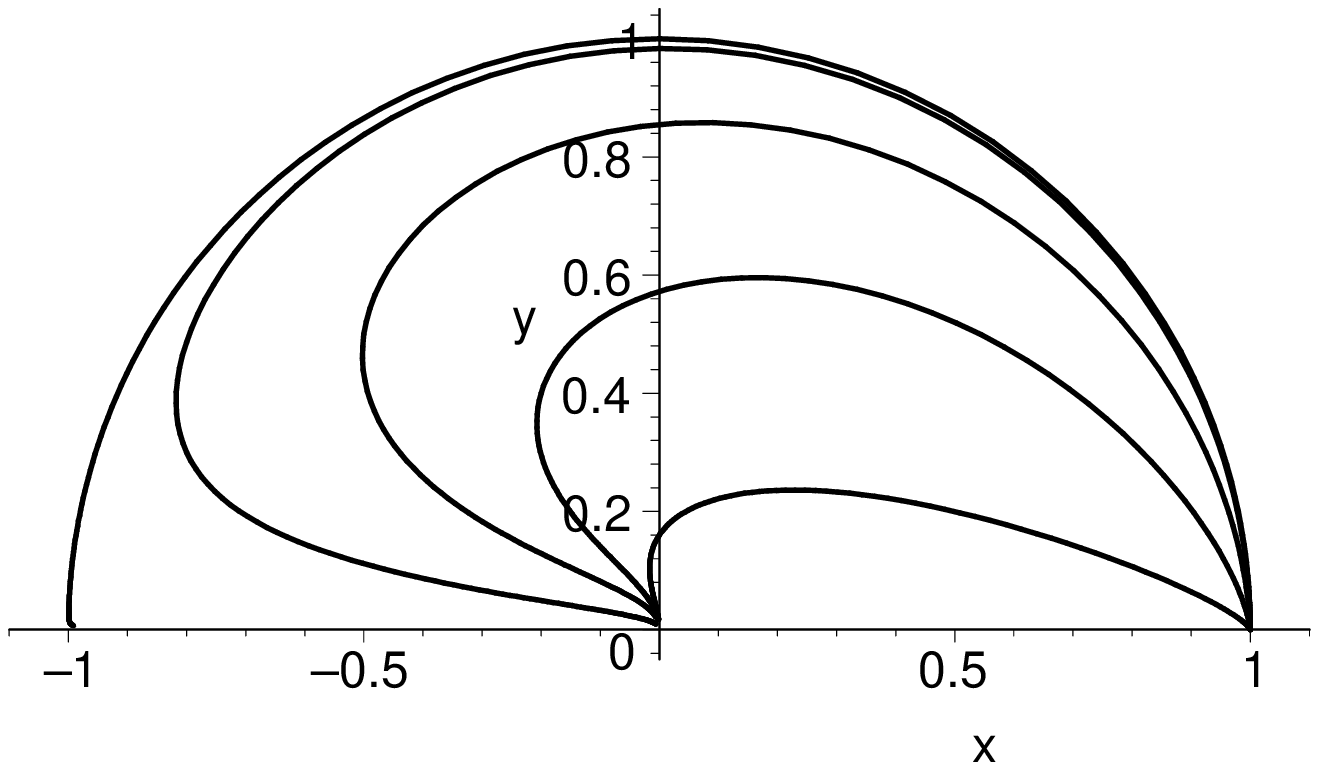}
  \includegraphics[width=5cm]{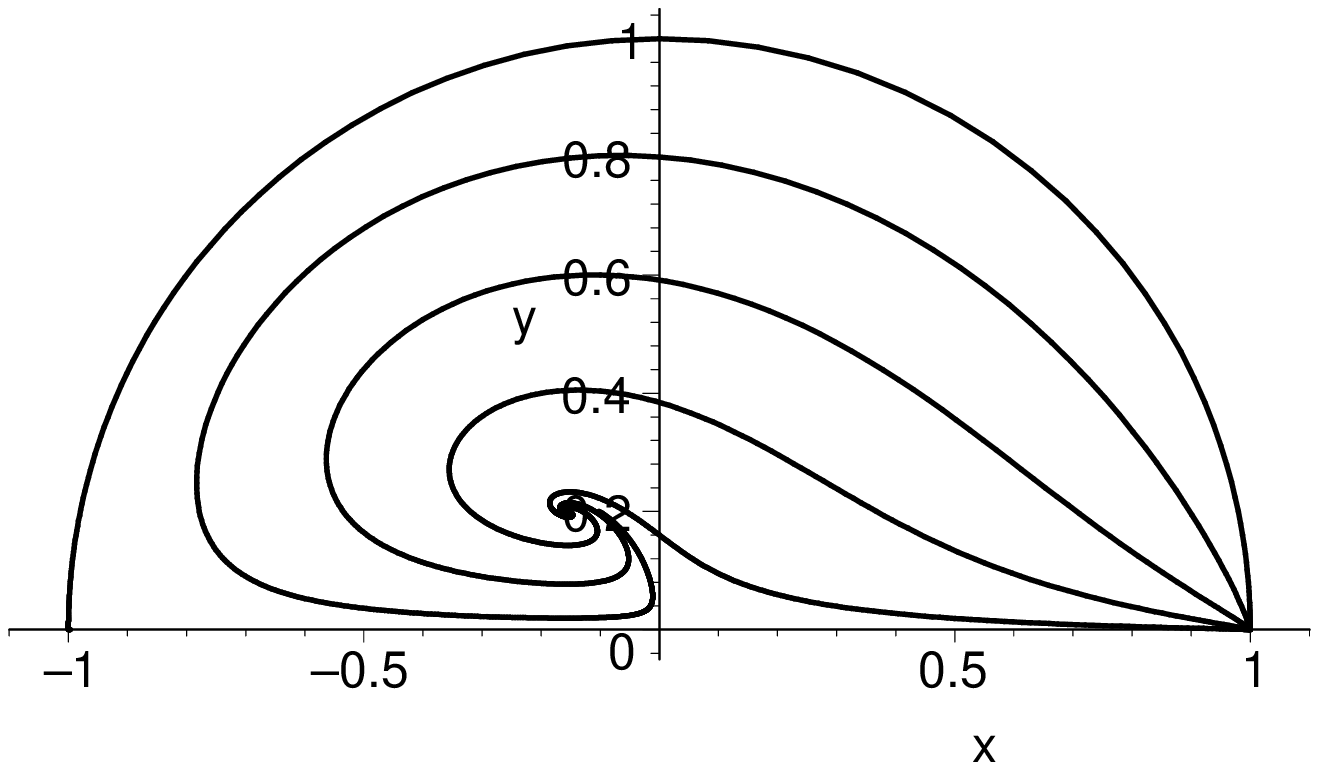}\\
\includegraphics[width=5cm]{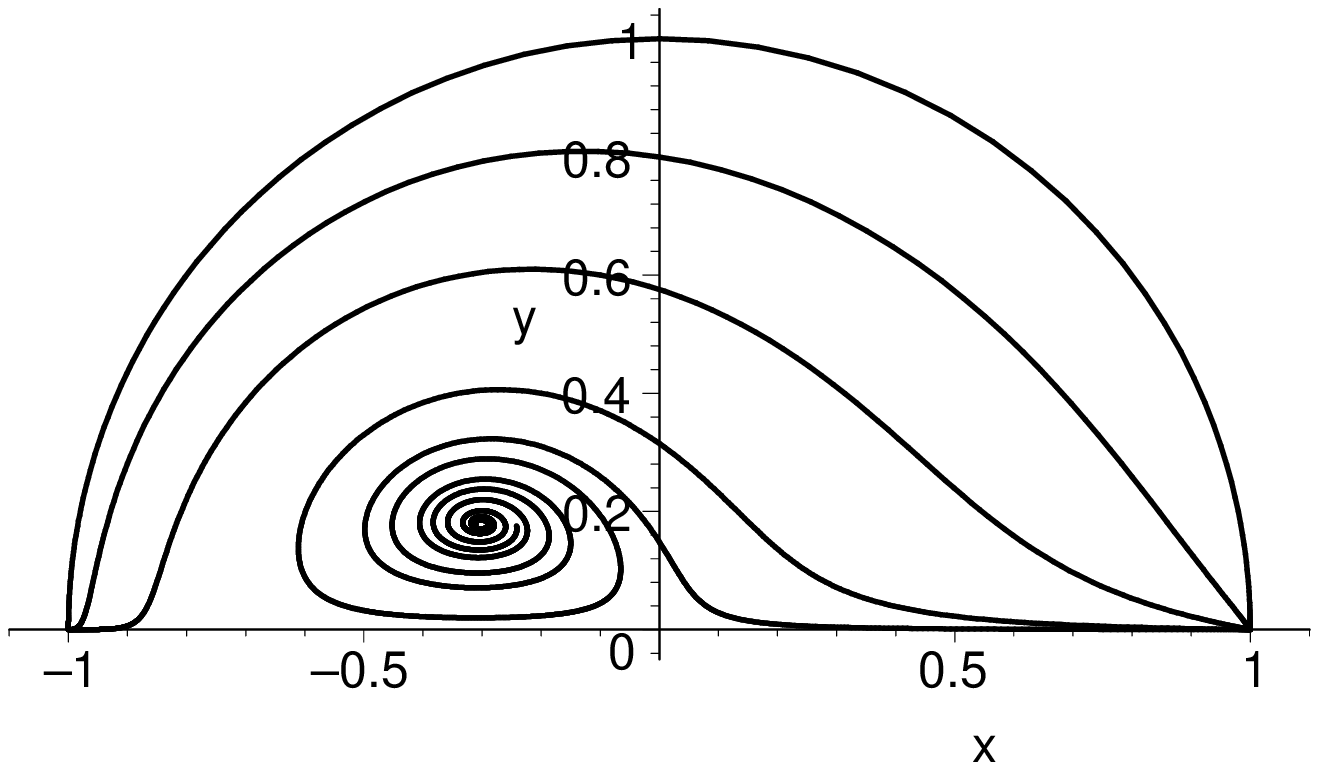}
\includegraphics[width=5cm]{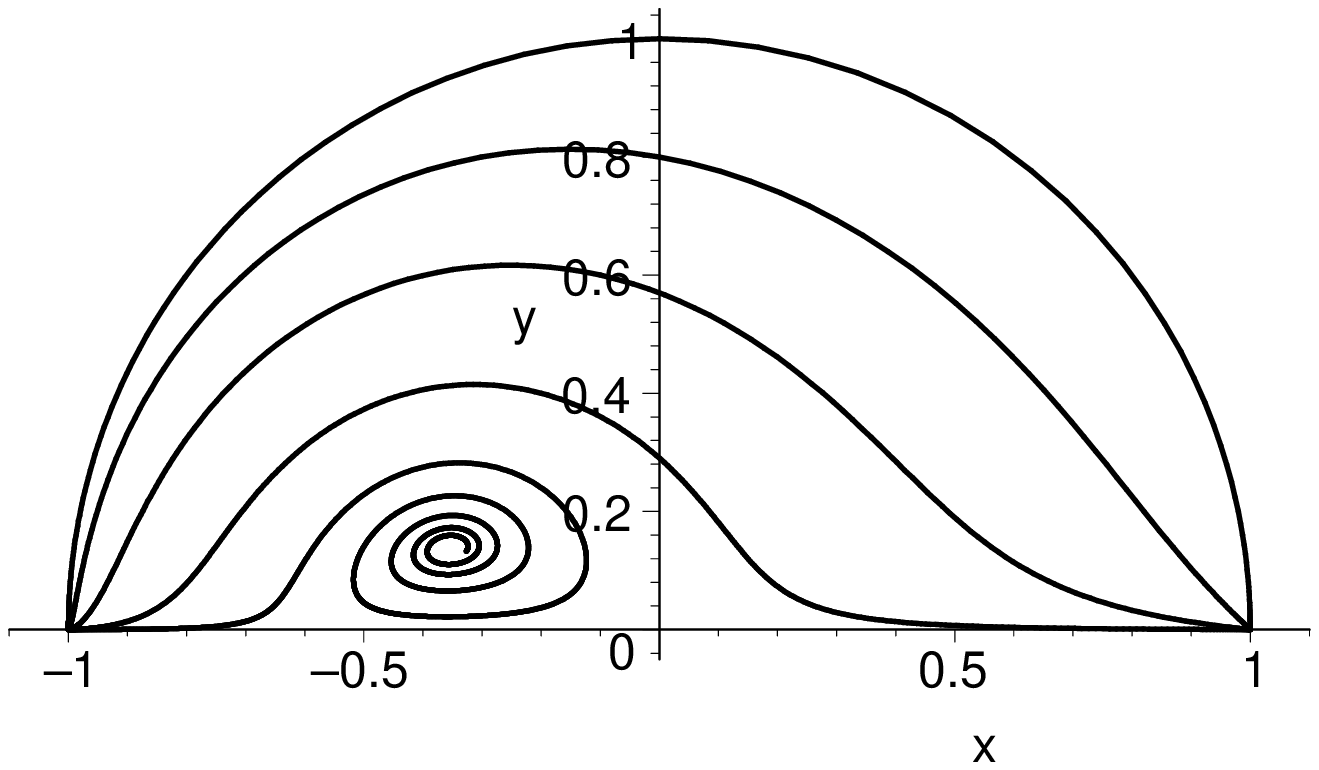}\\
\includegraphics[width=5cm]{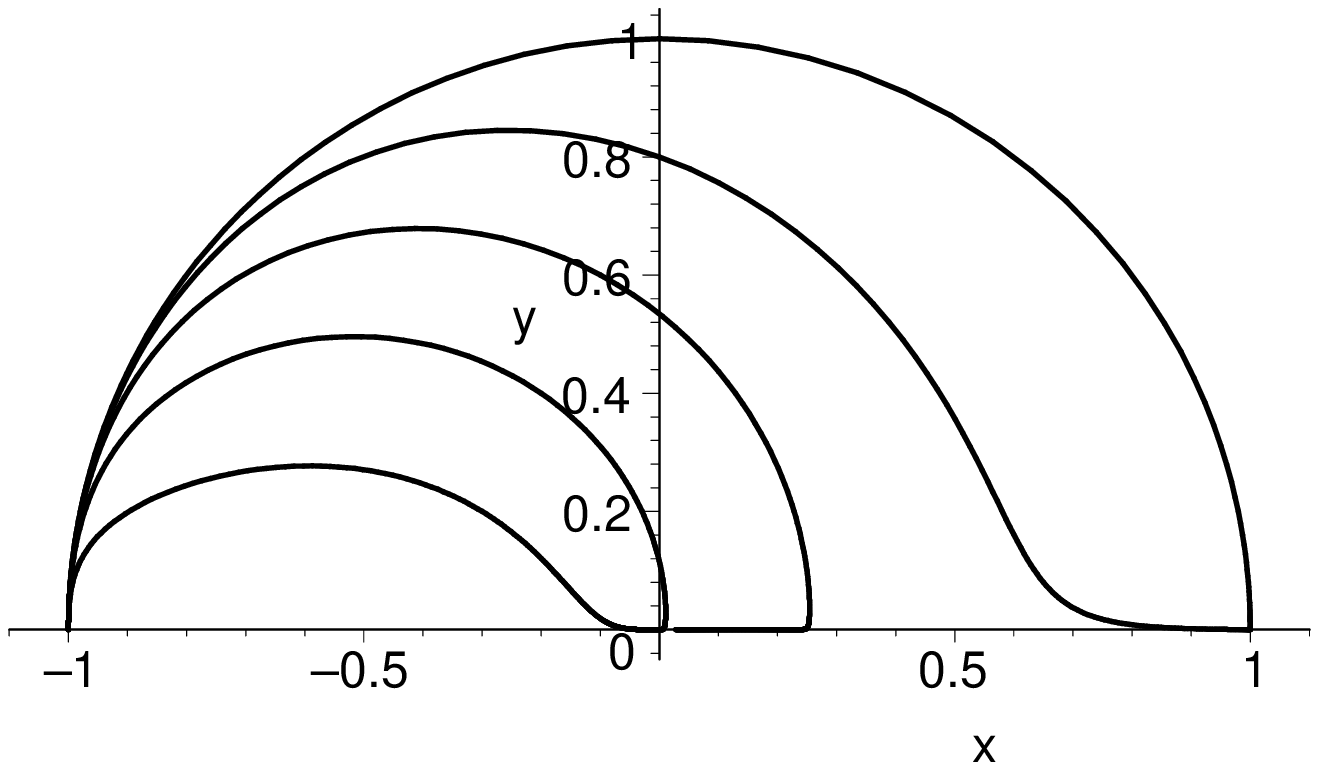}
\includegraphics[width=5cm]{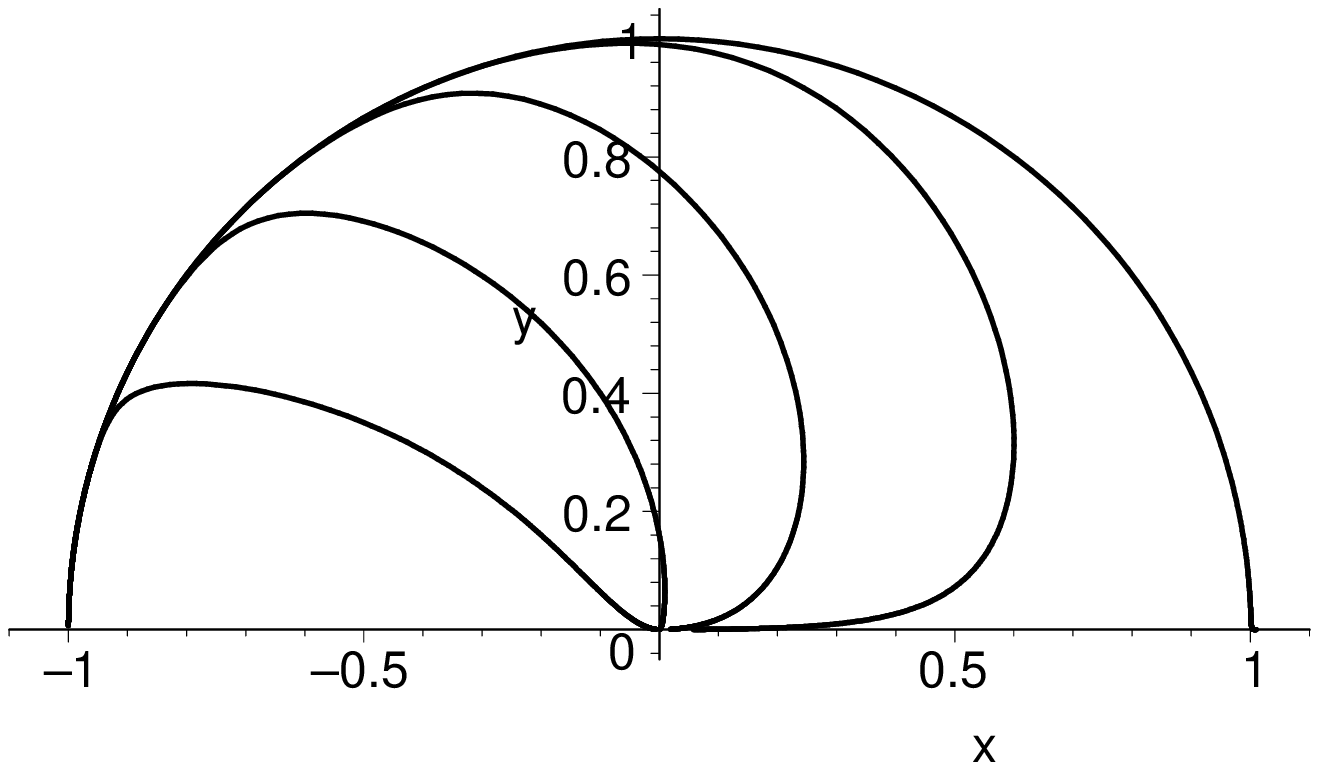}\\
  \caption{Solution curves and phase portraits for the dynamical system (\ref{eq:CFequations}). From top left to right: $\gamma=1.2$, $\gamma=1.4$, $\gamma=1.46$, $\gamma=1.48$, $\gamma=1.6$, and $\gamma=1.8$.}\label{Fig:CF}
\end{figure}

Based on the eigenvalues of the equilibrium points, the equilibrium points are future attractors in the range specified:
\begin{enumerate}
\item{}$\mathcal{CF}$: $10/9<\gamma\leq 4/3$.
\item{}$\widetilde{\mathcal{CF}}_{1\pm}$: $\frac 43<\gamma<\frac 56+\frac{\sqrt{721}}{42}$.
\item{} $\widetilde{\mathcal{CF}}_{2}$: Never an attractor.
\item{}$\widetilde{\mathcal{ECF}}_{-}$: $\frac{24-\sqrt{3}}{15}<\gamma<2$.
\end{enumerate}
In Figure \ref{Fig:CF} we have plotted some phase portraits for
different values of $\gamma$.

Including the previous analysis of the non-tilted equilibrium
points we can thus conclude:
\begin{itemize}
\item{} $2/3<\gamma<10/9$: Collins type VI$_{-1/9}$ perfect fluid solution is stable.
\item{} $\gamma=10/9$: Wainwright's type VI$^*_{-1/9}$, $\gamma=10/9$, solution \cite{Wainwright10-9} is stable.
\item{} $10/9<\gamma<2$: The Collinson-French vacuum is stable. Moreover, for $4/3<\gamma$ the asymptotic tilt velocity is non-zero and for $\frac 56+\frac{\sqrt{721}}{42}<\gamma$ it is extreme.
\end{itemize}

\section{Summary and Outlook}
In this paper we have considered tilted Bianchi models of solvable
type. We have used the dynamical systems approach and written down
the system of equations with one unspecified  gauge function. We
pointed out that various different gauge choices are appropriate
for different models and different applications. This formalism
was then used to study the late-time behaviour of various tilted
Bianchi models, with an emphasis on the existence of equilibrium
points and their stability properties.

\begin{table}
\centering
\begin{tabular}{|c|c|c|l|}
\hline
Bianchi &   &  &  \\
Type & Matter & Attractor & Comments \\
\hline \hline
I & $\frac 23<\gamma<2$ & $\mathcal{I}(I)$ & No tilt allowed  \\
\hline
II & $\frac 23<\gamma<\frac {10}7$ & $\mathcal{C}\mathcal{S}(II)$ & Non-tilted Collins-Stewart \\
 & $\frac{10}7<\gamma<\frac{14}9$  & $\mathcal{H}(II)$ & Hewitt's tilted type II \cite{Hewitt} \\
 & $\gamma=\frac{14}9$  & $\mathcal{L}(II)$ & Tilted type II bifurcation \\
 & $\frac{14}9<\gamma<2$  & $\mathcal{E}(II)$ & Extremely tilted \\
\hline
IV & $\frac 23<\gamma<2$ & Plane waves & Tilted/non-tilted  \\
\hline
V & $\frac 23<\gamma<2$ & Milne &  Tilted/non-tilted  \\
\hline
{}VI$_{h<-1}$ & $\frac 23<\gamma<\frac{2(1-h)}{1-3h}$ & $\mathcal{C}(VI_h)$ & Non-tilted Collins type VI$_h$ \\
  & $\frac{2(1-h)}{1-3h}<\gamma<2$ & Plane waves & Tilted/non-tilted  \\
\hline
VI$_{h>-1}$ & $\frac 23<\gamma<\frac{2(3+\sqrt{-h})}{5+3\sqrt{-h}}$ & $\mathcal{C}(VI_h)$ & Non-tilted Collins type VI$_h$ \\
  & $\frac{2(3+\sqrt{-h})}{5+3\sqrt{-h}} <\gamma<2$ & ? & No attractors known
\\
\hline
VI$^*_{-1/9}$ & $\frac 23<\gamma<\frac {10}9$ & $\mathcal{C}(VI_{-1/9})$ & Non-tilted Collins type VI$_{-1/9}$ \\
 & $\gamma=\frac{10}9$ & $\mathcal{W}(VI_{-1/9})$ & Non-tilted Wainwright type VI$_{-1/9}$ \\
 & $\frac{10}9<\gamma< 2$ & Collinson-French & Tilted/non-tilted \\
\hline
VI$_0$ & $\frac 23<\gamma<\frac 65$ & $\mathcal{C}(VI_{0})$ & Non-tilted Collins type VI$_0$ \\
& $\gamma=\frac 65$ & $\mathcal{L}(VI_0)$ & Tilted type VI$_0$ bifurcation \\
& $\frac 65<\gamma<2$ & $\mathcal{E}(VI_0)$ & Extremely tilted \\
\hline
VII$_h$ & $\frac 23<\gamma<2$ & Plane waves & Tilted/non-tilted  \\
\hline
VII$_0$ & $\frac 23<\gamma<2$ & ? & Analysis incomplete\\
\hline
\end{tabular}
\caption{The late-time behaviour of Bianchi  models with a tilted
$\gamma$-law perfect fluid (see the text for details and
references). The case $0<\gamma<2/3$ is covered by Corollary \ref{no-hair}.} \label{tab:outline}
\end{table}

Of particular interest are the plane-wave spacetimes of the class
B models.  In the analysis we have seen a new feature arise in the
dynamical behaviour of Bianchi cosmologies; namely, an oscillating
behaviour of the asymptotic tilt velocity. In the type IV model,
this oscillation occurs in a tiny region of the parameter space in
which the dynamical system undergoes a so-called Hopf-bifurcation,
and is related to the fact that there are no stable equilibrium
points for the corresponding range of parameters in the compact
state space. This Hopf-bifurcation is a manifestation of a more
general feature of the asymptotic behaviour of these models; their
extreme sensitivity to the parameter values. An outline of the
stability properties of the plane-waves was given and we pointed
out that they are future attractors in the set of type IV and
VII$_h$ tilted Bianchi models. These models are studied
further, both numerically and analytically, in a companion paper
\cite{HHC}.

The plane-wave analysis in the type IV models has emphasised that
studying the equilibrium points alone is not sufficient to
determine their late-time behaviour. This becomes even more
manifest for the type VII$_h$ models where more attracting closed
orbits appear. In addition,  numerical analysis
suggests that for the type VII$_h$ analogue of the loophole, the
solution curves will approach a compact surface which is
topologically a torus \cite{HHC}.
 As noted earlier, these interesting dynamical results
are not gauge effects; however, since the interesting mathematical
effects occur in the $\Omega=0$ invariant set, it is not clear
what the physical implications of this behaviour is.

 In a companion paper the tilted type IV
and VII$_h$ models are studied in greater detail, both
analytically and numerically \cite{HHC}. As in this paper, we will be mostly
interested in local behaviour close to equilibrium points,
although we will also consider some global aspects. In particular,
a more detailed analysis of the closed curves is performed. This analysis confirms the claims in this paper.

The late-time behaviour of tilted Bianchi models is summarized in
Table \ref{tab:outline}. This Table is based on the results in
this paper, and the following references: Type II \cite{HBWII};
Type V \cite{HWV}; Type VI$_h$ \cite{BHtilted}; Type
VI$^*_{-1/9}$ \cite{BHtilted}; Type VI$_0$ \cite{hervik}. As can
be seen from this table, there is a gap in our knowledge and it is
of interest to analyze the late-time behaviour of Bianchi type
VI$_h$ and type VII$_0$  models.  Both of these models seem
to have a different type of behaviour to that of the other Bianchi
models studied here; the type VI$_h$ model possibly has a matter-dominated
late-time behaviour and the type VII$_0$ model might have, in the
terminology of \cite{BHWeyl}, an \emph{extreme Weyl dominance}  at
late times. Here we have seen that this latter behaviour is also present in the one-component tilted type VII$_0$ subspace $T_1(VII_0)$. The remaining Bianchi models (which are excluded in
this work) are the two semi-simple models of type VIII and type IX.
Both of these models need a different approach than the one
adopted here.

We have focused on the late-time behaviour in this work. An
equally interesting task would be to consider the early time
behaviour. In \cite{HBWII} it was argued that the tilted type II
is \emph{chaotic} initially. We note that all Bianchi types,
except for type I and V, have type II as part of the boundary. For
the tilted type VI$_0$ model \cite{hervik} the type II boundary
played a crucial role in the initial singular behaviour, and since
the type II boundary is chaotic, it was argued that the tilted
type VI$_0$ is chaotic as well. We might therefore wonder whether
a similar phenomenon occurs in all the tilted Bianchi models
having type II as part of the boundary. Indeed, since type V does not have
the type II as part its boundary, we can ask whether tilted
Bianchi type V model is the only tilted Bianchi model
(apart from the trivial type I) which does not have chaotic
behaviour initially. A combined dynamical systems and metric approach \cite{Per} and a Hamiltonian analysis \cite{Jantzen} suggest that this is indeed the case. Moreover, there is some numerical evidence for this in the current tilted Bianchi models under consideration. We shall return to this question in future
work.

\section{Acknowledgments}
The authors would like to thank Robert van den Hoogen for discussions. 
This work was supported by NSERC (AC) and the Killam Trust and
AARMS (SH).

\appendix
\section{Consistency of the constraints}
We define the following functions:
\beq
H&=& \Sigma^2+A^2+|{\bf N}_{\times}|^2+\Omega-1\nonumber \\
Q_1&=& 2\Sigma_+A+2\mathrm{Im}({\mbold\Sigma}_{\times}^*{\bf N}_{\times})+\frac{\gamma\Omega v_1}{G_+} \nonumber \\
{\bf Q} &=&
\mbold{\Sigma}_{1}(i\bar{N}-\sqrt{3}A)+i{\mbold{\Sigma}}^*_{1}{\bf
N}_{\times}+\frac{\gamma\Omega {\bf v}}{G_+} \nonumber
 \\
G &=& A^2+3h\left(|{\bf N}_{\times}|^2-\bar{N}^2\right). \eeq The
constraints are satisfied if and only if all of these functions
vanish. To check the consistency of these constraints, we assume
first that they do not vanish and calculate their
time-derivatives. After some algebra we obtain \beq
H'&=& 2qH+2AQ_1, \nonumber \\
Q_1'&=& 2(q-1+\Sigma_+)Q_1-2\sqrt{3}\mathrm{Re}\left({\mbold\Sigma}_1{\bf Q}^*\right), \nonumber \\
{\bf Q}'&=& (2q-2-\Sigma_++i\phi'){\bf Q}-\sqrt{3}{\mbold\Sigma}_{\times}{\bf Q}^*, \nonumber \\
G'&=& 2(q+2\Sigma_+)G.
\eeq
This means that if all the constraints vanish initially, they will be zero at all times.

\section{Some proofs}
\subsection{Outline of the proof of Theorem \ref{thm:V+VIIh}}
\label{App:Proof} We will prove Theorem \ref{thm:V+VIIh} for the
Bianchi type VII$_h$ models by contradiction. We assume, therefore,
that there exists an equilibrium point with ${\bf
v}\neq 0$ and  $\bar{N}^2-\left|{\bf N}_{\times}\right|^2>0$ and
$A\neq 0$. The outline of the proof goes as follows.

We first assume that ${\bf N}_{\times}=0$. This leads to
\[ \bar{N}\neq 0~ \Rightarrow ~ {\mbold\Sigma}_{\times}=0~\Rightarrow~ {\mbold\Sigma}_1^2=-\frac{\gamma\Omega}{2G_+}{\bf v}^2.\]
Using the constraint equation we see that $\Omega\neq 0$  and
$(i\bar{N}-\sqrt{3}A)^2$ equals a negative real number. Thus
$A=0$, which is a contradiction. We can thus assume  that ${\bf
N}_{\times}\neq 0$. This allows us to use the N-gauge, and we will
assume this gauge is chosen henceforth.

Let us now assume that $\Omega\neq 0$. Using the evolution equations for
$A$, ${\bf N}_{\times}$, ${\mbold\Sigma}_1$, and ${\bf v}$ we obtain
$\Sigma_{23}=0$, $T=\mathcal{S}$, $q=2+3\mathcal{S}$,
$Av_1=-1-(5/2)\mathcal{S}$, $\Sigma_+=-1-(3/2)\mathcal{S}$, and
$\Sigma_-=v_1N_{23}$. Inserting this information into the $\Omega$ equation we
 obtain  the two separate cases: (i)
$V^2=1$ , and (ii) $\gamma=6/5$. After some lengthy calculations
both of these cases lead to contradictions. These calculations
involve solving the remaining equations and making sure that all
parameters are real and obey the required bounds; i.e.,
$0<\Omega<1$, $0<V^2\leq 1$, etc.

Therefore, we must have $\Omega=0$. In this case the equations of
motion decouple and we can use the non-tilted analysis which implies
that the only equilibrium points are the plane-wave spacetime.
From eq.(\ref{eq:varphiVIIh}) we require  $\varphi'=0$, which implies $x=1$ ($\lambda^2>1$). Since $x=v_1$, 
this implies that $V^2=v_1^2+v_2^2+v_3^2\geq 1$. However, since $V^2\leq 1$, we must have $V^2=1$ and $v_2=v_3=0$ which is a contradiction. 
Since ${\bf v}=0$ implies ${\mbold\Sigma}_1=0$, Theorem \ref{thm:V+VIIh} now follows. 

\subsection{Proof of Theorem \ref{thm:VII0>rad}}
\label{proof:VII0}
The Proof consists of two steps: \textit{(i)}: The existence of a monotonic function. \textit{(ii)}: 
Showing that the monotonic function forces $\Sigma_+\rightarrow -(3\gamma-4)/2$ at late times.

The first step is to find a monotonic function for the reduced system and amend it so that it is monotonic for the full system at late times. Such a function is:
\beq
\hat{Z}_4=\frac{\sigma_1^m\beta\Omega}{(1+m\Sigma_+)^{2(1+m)}}, \quad m\equiv\frac 12(3\gamma-4),
\eeq
where $\beta$ is given in eq.(\ref{def:beta}). We amend this function by defining\[ \bar{Z}_4=\frac{\hat{Z}_4}{1+fM\cos\psi}, \quad f\equiv\frac{\rho\left[\sigma_1(1-m^2)-m(\Sigma_++1)(1+m\Sigma_+)\right]}{2\sqrt{3}\sigma_1(1+m\Sigma_+)}. \] 
Note that since $|\rho/\sigma_1|\leq 1$ and  $m\leq 1$, $f$ is bounded. At sufficiently late times we have
\beq
\frac{\bar{Z}_4'}{\bar{Z}_4}=\frac{2(\Sigma_++m)^2}{(1+m\Sigma_+)}+MB,
\eeq
where $B$ is a function of the state space variables and is bounded. Thus at late times, $\bar{Z}_4$ is monotonically increasing. However, for $4/3<\gamma<2$, this function is bounded for sufficiently late times, and hence, we get the bound 
\beq
\int_{\tau_0}^\infty (\Sigma_++m)^2\mathrm{d}\tau<\infty, 
\label{eq:intsigma}\eeq
for a sufficiently large $\tau_0$. 

The second step is to show that this implies $\Sigma_+\rightarrow -m$. We will show this by contradiction. 

Define the everywhere positive function $f(\tau)\equiv(\Sigma_++m)^2$. Assume that $f\nrightarrow 0$. Then 
there must exist a $\delta>0$ and an infinite sequence $\{\tau_n\}$ with $\lim_{n\rightarrow\infty}\tau_n=\infty$
such that
\beq
f(\tau_n)\geq \delta, \quad \forall n. 
\eeq
From the evolution equations we note that $f'(\tau)$ is bounded; i.e. there exists a $D>0$ such that $|f'(\tau)|\leq D$ for all $\tau$. 
We then define the family of functions $g_n$ given by 
\beq
g_n(\tau)\equiv\begin{cases} 
\delta -D|\tau_n-\tau|, & \tau_n-\delta/D\leq \tau\leq \tau_n+\delta/D \\
0, & \text{otherwise}. 
\end{cases} 
\eeq
We note that $0\leq g_n(\tau)\leq f(\tau)$, and $\int_{\tau_0}^{\infty}g_n(\tau)\mathrm{d}\tau=\delta^2/D$. Moreover, by defining (we can without loss of generality 
assume that $\mathrm{support}(g_i)\cap\mathrm{support}(g_j)=\emptyset$ for $i\neq j$)
\[ g(\tau)\equiv \sum^{\infty}_n g_n(\tau), \] 
we have $0\leq g(\tau)\leq f(\tau)$. Thus
\beq
\int_{\tau_0}^{\infty}f(\tau){\mathrm d}\tau\geq \int_{\tau_0}^{\infty}g(\tau){\mathrm d}\tau=\sum_n^{\infty}\int_{\tau_0}^{\infty}g_n(\tau){\mathrm d}\tau=\frac{\delta^2}{D}\sum_n^{\infty}1=\infty, 
\eeq
which contradicts eq.(\ref{eq:intsigma}). Hence, $\Sigma_+\rightarrow -m$ which proves the Theorem.

\section{Eigenvalues for non-exceptional types.}
Below, we have defined $\tilde{h}=1/\sqrt{|h|}$ and
\beq
\beta=\begin{cases}
i\tilde{h}, & h>0, \\
0, & \text{IV}, \\
\tilde{h}, & h<0.
\end{cases}
\eeq
\subsection{Plane waves, $\mathcal{L}(h)$ (for all $h$):}
\beq
&& \lambda_1=0, \quad \lambda_{2,3}=-2\left[(1+\Sigma_+)\pm 2i\sqrt{3}N_{23}\right], \quad \lambda_4=-4\Sigma_+-(3\gamma-2), \nonumber \\
&& \lambda_5=(3\gamma-4)+2\Sigma_+, \quad \lambda_{6,7}=(3\gamma-4)-\Sigma_+\pm (1+\Sigma_+) \beta.\nonumber
\eeq
\subsection{Plane waves, $\tilde{\mathcal{L}}(h)$ (for all $h$):}
\beq
&& \lambda_{1}=0, \quad \lambda_{2,3}=-2\left[(1+\Sigma_+)\pm 2i\sqrt{3}N_{23}\right], \quad \lambda_4=-\frac{2-\gamma}{\gamma-1}(1-2\Sigma_+),  \nonumber \\ &&
\lambda_5=-\frac{(1-2\Sigma_+)(5\gamma-6+2\gamma\Sigma_+)(3\gamma-4+2\Sigma_+)}{(\gamma-1)(9\gamma-10+4\Sigma_+(1-\Sigma_+))}, \nonumber \\ &&
 \lambda_{6,7}=\frac{(1-2\Sigma_+)(3\gamma-4)}{2(\gamma-1)}\pm (1-v_1)(1+\Sigma_+)\beta.\nonumber
\eeq
\subsection{Plane waves, $\tilde{\mathcal{L}}_+(h)$ (for all $h$):}
\beq &&
\lambda_{1}=0, \quad \lambda_{2,3}=-2\left[(1+\Sigma_+)\pm 2i\sqrt{3}N_{23}\right], \quad \lambda_4=0,   \nonumber\\
&& \lambda_{5}=2(1-2\Sigma_+), \quad \lambda_{6,7}=1-2\Sigma_+.
\eeq
\subsection{Plane waves, $\tilde{\mathcal{L}}_-(h)$ (for all $h$):}
\beq &&
\lambda_{1}=0, \quad \lambda_{2,3}=-2\left[(1+\Sigma_+)\pm 2i\sqrt{3}N_{23}\right], \quad \lambda_4=-4(1+\Sigma_+),   \nonumber\\
&& \lambda_{5}=-\frac{2(5\gamma-6+2\gamma\Sigma_+)}{2-\gamma}, \quad \lambda_{6,7}=-1-4\Sigma_+\pm 2(1+\Sigma_+)\beta.\nonumber
\eeq
\subsection{Plane waves, $\tilde{\mathcal{F}}_{\pm}(IV)$:}
\beq &&
\lambda_{1}=0, \quad \lambda_{2,3}=-2\left[(1+\Sigma_+)\pm 2i\sqrt{3}N_{23}\right], \quad \lambda_4=-\frac{(1-2\Sigma_+)(5\gamma-6)}{3-2\gamma},   \nonumber\\
&& \lambda_{5}=0, \quad {\lambda_{6}\lambda_{7}}=-\frac{(1-2\Sigma_+)^2(4-3\gamma)(3\gamma-4-\Sigma_+)(3-2\gamma+\gamma\Sigma_+)}{(3-2\gamma)G(\gamma,\Sigma_+)}\nonumber  \nonumber \\
&& \lambda_6+\lambda_7=\frac{(1-2\Sigma_+)F(\gamma,\Sigma_+)} {4(3-2\gamma)(17\gamma^2-40\gamma+24)G(\gamma,\Sigma_+)}
\nonumber \eeq
where
\beq
G(\gamma,\Sigma_+)&\equiv & (5\gamma-6)\Sigma_+^2-(18-25\gamma+9\gamma^2)\Sigma_+-3+2\gamma, \nonumber \\
F(\gamma,\Sigma_+)&\equiv & \left[2(17\gamma^2-40\gamma+24)\Sigma_+-33\gamma^3+121\gamma^2-152\gamma+66\right]^2\nonumber \\&&
- 9(\gamma-1)^2\left(121\gamma^4-736\gamma^3+1664\gamma^2-1656\gamma+612\right)
\label{def:Fdef}\eeq
Here is $G(\gamma,\Sigma_+)<0$ in the whole region under consideration. $F(\gamma,\Sigma_+)=0$, defines a line from $(4/3,0)$ to $(4/3,-1/4)$.

\subsection{Plane waves, $\tilde{\mathcal{E}}_{\pm}(IV)$:}
\beq &&
\lambda_{1}=0, \quad \lambda_{2,3}=-2\left[(1+\Sigma_+)\pm 2i\sqrt{3}N_{23}\right], \quad \lambda_4=\frac{(1-2\Sigma_+)(1+2\Sigma_+)}{\Sigma_+},   \nonumber\\
&& \lambda_{5}=0, \quad {\lambda_{6}}=-\frac{(1-2\Sigma_+)(1+4\Sigma_+)}{3\Sigma_+}, \quad {\lambda_{7}}=\frac{2(1-2\Sigma_+)(2\gamma-3-\gamma\Sigma_+)}{3\Sigma_+(2-\gamma)}, \nonumber
\eeq
\subsection{Plane waves, $\tilde{\mathcal{F}}_{\pm}(h)$:}
\beq&&
\lambda_{1}=0, \quad \lambda_{2,3}=-2\left[(1+\Sigma_+)\pm 2i\sqrt{3}N_{23}\right], \nonumber \\ && \lambda_4=-\frac{(1-2\Sigma_+)[(5\gamma-6)\pm \tilde{h}(2-\gamma)]}{3-2\gamma\pm \tilde{h}},  \quad
\lambda_{5}=2(1-v_1)\alpha, \nonumber \\ && {\lambda_{6}\lambda_{7}}\propto(1-2\Sigma_+)^2[(4-3\gamma)(1+\Sigma_+)+\alpha(2-\gamma)]\nonumber \\ && \qquad \quad \times (3\gamma-4-\Sigma_+-\alpha)[3-2\gamma+\gamma(\Sigma_++\alpha)]\nonumber  \\
&& \lambda_6+\lambda_7=\text{something really awful}
\eeq
where $\alpha\equiv \pm(1+\Sigma_+)\tilde{h}.$
The expressions for $\lambda_6$ and $\lambda_7$ are too nasty to write down, but $\lambda_6+\lambda_7=0$ defines a curve which ultimately rives rise to the type VI$_h$ loophole.

\subsection{Plane waves, $\tilde{\mathcal{E}}_{\pm}(h)$:}
\beq&&
\lambda_{1}=0, \quad \lambda_{2,3}=-2\left[(1+\Sigma_+)\pm 2i\sqrt{3}N_{23}\right], \quad \lambda_4=\frac{(1-2\Sigma_+)(3+6\Sigma_++2\alpha)}{3\Sigma_++\alpha},   \nonumber\\ &&
\lambda_{5}=2(1-v_1)\alpha, \quad {\lambda_{6}}=-\frac{(1-2\Sigma_+)(1+4\Sigma_++2\alpha)}{3\Sigma_++\alpha}, \nonumber \\ && {\lambda_{7}}=\frac{2(1-2\Sigma_+)[2\gamma-3-\gamma(\Sigma_++\alpha)]}{(3\Sigma_++\alpha)(2-\gamma)}, \nonumber
\eeq
where $\alpha\equiv \pm(1+\Sigma_+)\tilde{h}.$
\section{Eigenvalues for the Collinson-French vacuum}
Below are listed the eigenvalues corresponding to the eq. (\ref{eq:CFequations}), $\lambda_{1,2}$, and the eigenvalue corresponding to the matter density, $\Omega$, which is denoted $\lambda_3$.
\subsection{Collinson-French, $\mathcal{CF}$:}
\beq
\lambda_1=\frac 13(9\gamma-14),\quad \lambda_2=(3\gamma-4), \quad \lambda_3=-\frac 13(9\gamma-10).\nonumber
\eeq
\subsection{Collinson-French, $\widetilde{\mathcal{CF}}_{1\pm}$:}
\beq
\lambda_{1,2}=\frac{s\pm\sqrt{s^2-4d}}{2}, \quad \lambda_3=-\frac{10\gamma(4\gamma^2-9\gamma+6)}{3(3-\gamma)(3\gamma^2-9\gamma+10)}\nonumber
\eeq
where
\beq
d\equiv\lambda_1\lambda_2=\frac{100(3\gamma-4)(3-2\gamma)}{3(3-\gamma)(3\gamma+1)}, \quad s\equiv \lambda_1+\lambda_2=\frac{105\left[\left(\gamma-\frac 56\right)^2-\frac{721}{42^2}\right]}{3(3-\gamma)(3\gamma+1)}\nonumber.
\eeq

\subsection{Collinson-French, $\widetilde{\mathcal{CF}}_{2}$:}
\beq
&&\lambda_1=-\frac{10(3-2\gamma)}{3(\gamma-1)},\quad \lambda_2=\frac{(9\gamma-14)\left[(9\gamma-14)^2-6(\gamma-1)^2\right]}{3(81\gamma-258\gamma+202)},\nonumber \\
&&\lambda_3=-\frac{10\gamma(12\gamma^2-39\gamma+32)}{3(\gamma-1)(81\gamma^2-246\gamma+190)}.\nonumber \\
\eeq
\subsection{Collinson-French, $\widetilde{\mathcal{ECF}}_{\pm}$:}
\beq
\lambda_{1}=\frac 23\left(1\pm\sqrt{6}\right), \quad \lambda_2=-\frac{2\left[9\gamma-14\mp\sqrt{6}(\gamma-1)\right]}{3(2-\gamma)}, \quad \lambda_3=-\frac 13(4\mp\sqrt{6}).\nonumber
\eeq
\bibliographystyle{amsplain}

\end{document}